\documentclass[universe,review,accept,oneauthor,pdftex]{mdpi}
\usepackage{color}
\usepackage{xcolor}
\usepackage[normalem]{ulem}
\graphicspath{{./Figures/}}
\firstpage{1} 
\pubvolume{8}
\issuenum{2}
\articlenumber{95}
\pubyear{2022}
\copyrightyear{2022}
\externaleditor{Academic Editors: {Chitta Ranjan Das, Alexander S. Barabash and Giuseppe~Verde}} %MDPI: please add Academic Editor.
\datereceived{6 December 2021} 
\dateaccepted{31 January 2022} 
\datepublished{3 February 2022} 
\hreflink{https://doi.org/10.3390/\linebreak universe8020095} % If needed use \linebreak
\newcommand{\cd}{\makebox[0.08cm]{$\cdot$}}
\usepackage{bm}
\Title{Abnormal Bound Systems}

\TitleCitation{{Abnormal Bound Systems}} % Please confirm

\Author{{Vladimir A.~Karmanov~\href{https://orcid.org/0000-0001-9347-5352}{\orcidicon}}} %MDPI: 1. Please carefully check the accuracy of names and affiliations.  2. please add the whole name

\AuthorNames{Vladimir A.~Karmanov}

\AuthorCitation{Karmanov, V.A.}
\address[1]{%
Lebedev Physical Institute, Leninsky Prospect
53, 119991 Moscow, Russia; karmanovva@lebedev.ru}%; karmanovva@lebedev.ru}

\abstract{It is taken for granted that  bound systems are made of massive constituents that interact through particle exchanges (charged particles interacting via photon exchanges, quarks in elementary particles interacting via gluon exchanges, and nucleons in nuclei interacting via meson exchanges). However, as   was recently  theoretically found, there exist 
 systems dominated  by  exchange particles (at least for the zero exchange masses). In these systems, the contribution of massive constituents is negligible.
These systems have a relativistic nature (since they are mainly made of   massless particles moving at the speed of light), and therefore, they cannot be described by the Schr\"odinger equation. 
Though these results were found so far in the simple Wick--Cutkosky model (spinless constituents interacting via the ladder of spinless massless  exchanges), the physical ground for  their existence seems to be rather general.} % Please confirm that meaning has been retained


\begin{document}
%%%%%%%%%%%%%%%%%%%%%%%%%%%%%%%%%%%%%%%%%%

 %\setcounter{section}{-1} %% Remove this when starting to work on the template.
 %\section{How to Use this Template}

 %The template details the sections that can be used in a manuscript. Note that the order and names of article sections may differ from the requirements of the  %journal (e.g., the positioning of the Materials and Methods section). Please check the instructions on the authors' page of the journal to verify the correct  %order and names. For any questions, please contact the editorial office of the journal or support@mdpi.com. For LaTeX-related questions please contact  %latex@mdpi.com.

%The order of the section titles is: Introduction, Materials and Methods, Results, Discussion, Conclusions for these journals: aerospace,algorithms,antibodies,antioxidants,atmosphere,axioms,biomedicines,carbon,crystals,designs,diagnostics,environments,fermentation,fluids,forests,fractalfract,informatics,information,inventions,jfmk,jrfm,lubricants,neonatalscreening,neuroglia,particles,pharmaceutics,polymers,processes,technologies,viruses,vision

\section{Introduction}

Bound states have an essentially non-perturbative nature. They appear\; if\; the\; coupling constant exceeds
some critical value (for an interaction of a finite radius). Their binding energies (and, therefore, the wave functions) cannot be calculated perturbatively. Let us consider, for example, a two-body system interacting with the Coulomb potential $V(r) =-\frac{\alpha}{r}$. Though, in this potential, the coupling constant $\alpha$ can take any small value 
(the bound states exist for any small  $\alpha$), the binding energy vs. $\alpha$ is quadratic: $E_n=-\frac{m\alpha^2}{4n^2}$, which already demonstrates its non-perturbative nature. If the  perturbative 
%first order 
correction to the free state in terms of the potential would be valid, in the first order, it should be linear vs. $\alpha$. A non-perturbative solution implies that the scattering amplitude near the bound-state pole and binding energy are determined, in general, by the sum of infinite series in $\alpha$, i.e., by an infinite number of exchanges. Then, a natural question arises: Why, in view of the infinite number of exchanges, are we dealing with a system containing massive constituents, not with a system  containing, in addition to massive constituents, an indefinite (or even infinite) number of massless exchange particles? 

We claim that the systems of both types are predicted by theory \cite{cks21}. However, so far, we dealt mainly with non-relativistic systems, in which 
the effects of retardation  in the interaction are not important. It is implied that the interaction is  instantaneous. For this interaction, the lifetime of the virtual exchange particles in the system is zero (they are absorbed immediately after emission). Therefore, in its intermediate state,  the system is dominated by slow (non-relativistic) particles. With  the  interaction approximated by a static potential, a few-body system is described by the Schr\"odinger equation. In this way, 
the intermediate states with many-body exchange particles are cut from the very beginning. This is a good non-relativistic approximation. 
%The exchange particles can be taken into account perturbatively, as a small correction.

On the other hand, if systems of the second type (with many massless exchange particles in the intermediate states) exist, they should be described by a relativistic equation. Moreover, since a relativistic approach covers the full domain of momenta, both small and large,  it can be applied to the systems of both types simultaneously: ({\it i}) the non-relativistic ones, also described  by the Schr\"odinger equation---the limiting case of the initial relativistic equation (the hydrogen atom in our case)---and ({\it ii}) purely relativistic systems, dominated by exchange particles, which cannot be discovered in the  Schr\"odinger framework. 

These qualitative considerations were recently confirmed by scrutinizing solutions of the relativistic Bethe--Salpeter (BS) equation. {Though the story started 
almost 70 years ago with Wick and Cutkosky}  \cite{wick,cutk} solving the BS equation \cite{bs}  {for the spinless constituents interacting via the ladder of spinless massless  exchanges, two types of solutions have been~found.} % Please confirm that meaning has been retained 

One of them reproduced the normal Coulomb spectrum and wave function in the non-relativistic limit.  The second one had no  non-relativistic counterpart. It was absent in the Schr\"odinger equation, it was found in the BS equation, and it disappeared in the non-relativistic limit. Therefore, it was called ``abnormal''. Its origin confused researchers and its nature was discussed in the literature over a couple of decades (see, e.g., \S 8 in \cite{nak69} and references therein). In general, two irreconcilable hypotheses were put forward.  ({\it i}) The abnormal solutions are a mathematical oddity of the BS equation. They have no  physical sense. No physical system can be mapped to them. ({\it ii})  The abnormal solutions correspond to physical systems. They at least contribute to the S-matrix. 

A breakthrough in understanding the abnormal solutions  occurred after analyzing their content.
% from point of view of composition of the Fock sectors of the state vector.
 It was found  \cite{cks21} that the valence contribution (of the massive constituents) is small and tends to zero when the binding energy decreases. Therefore, the abnormal states are dominated by the exchange particles. They can be called ``hybrid systems'', as they contain a few constituents (two in the  
Wick--Cutkosky model considered below)  and many exchange particles.
 This clarifies  their nature.
%However, the BS amplitude (\ref{bs}) is not enough to extract the full information about the system and this distribution.

{The present paper is devoted to a review of these systems, which are dominated by  massless exchange particles, as well as of their origins and properties.} % Please confirm that meaning has been retained
 It is based on the results published in Ref. \cite{cks21}, as well as in our reports  \cite{abnorm,dubna2018,LC2019_1} at conferences. In Section \ref{sv}, in the presentation of the field-theoretical background, we will not enter into the details of the formalized scheme. We will rely on a qualitative physical picture and designate only the outline. In the following sections, the presentation will be more precise. In Section \ref{bsa}, definition of the BS amplitude and its relation with the two-body wave function and the BS equation will be given.  The method  of solving the BS equation proposed by Wick and Cutkosky \protect{\cite{wick,cutk}} is explained in Section  \protect{\ref{WCm}}. The origin of extra solutions of the BS equation found with this method
(which are absent in the non-relativistic approach) is clearly demonstrated in Section  \protect{\ref{WCm}}.
 The two-body contributions to the full normalization of the state vector for the solutions of different natures are calculated in Section   \ref{2b}. The elastic and transition electromagnetic form factors for different solutions (both normal and abnormal) are presented in Section \ref{ffs}. The results are discussed in Section \ref{discuss}.
Some technical details of calculating the full norm of the state vector are included in Appendix \ref{app1}.

%%%%%%%%%%%%%%%%%%%%%%%%%%%%%%%%%
\section{The State Vector}\label{sv}

It should be clarified that {when we refer to the content  of a  physical system and its
 constituents,} we imply that any system is described by the field-theoretical state vector 
$|p\rangle$.  % Please confirm that meaning has been retained
This state vector is an eigenstate of the Hamiltonian $H=H_0+H_{int}$, where $H_0$ is the free Hamiltonian describing the free constituent and exchange fields, and
 $H_{int}$ is the interaction between them.  The four-momentum $p$ is the momentum of  the entire system. The eigenstates of the free Hamiltonian $H_0$ are the states $|n\rangle$ with a given number of  the free constituent and exchange particles.  The state vector  $|p\rangle$ can be decomposed in terms of these free states $|n\rangle$ (as the bound-state wave function can be decomposed in the plane waves---the Fourier transform); each of them corresponds to a fixed (and different) number $n$. Schematically:
\begin{equation}\label{p}
|p\rangle=\sum_{n}^{\infty}\psi_n |n\rangle.
\end{equation}

A superposition of the states with different numbers of particles appears, since the interaction $H_{int}$ does not conserve the number of particles. 
The state $|n\rangle$, which contains both ``constituent'' (valence) and ``exchange'' particles---a fixed number of them---is called the ``Fock sector'', whereas the coefficients  $\psi_n$ of this decomposition, which correspond not only   
to a fixed number of particles, but also to their fixed momenta, are called the ``Fock components''. According to the general rules of quantum mechanics, they determine the probabilities of finding  $n$ particles with  momentum distributions of 
$|\psi_n|^2\equiv |\psi_n(\vec{k}_1,\vec{k}_2,\ldots, \vec{k}_n;\vec{p})|^2$ in the system described by the state vector $|p\rangle$. The integral over the momenta  $\vec{k}_1,\vec{k}_2,\ldots, \vec{k}_n$
 determines the probability $N_n(\vec{p})$ of finding   $n$ particles in the system.
 % (say, 2 constituents and $n-2$ exchange particles), 
 The sum over $n$ is normalized to 1:  
 \begin{equation}\label{normstv}
 \langle p | p\rangle=\sum_n N_n(\vec{p})=1. 
 \end{equation}
 
In a non-relativistic two-body system, the probability $N_2$ of finding two constituent particles dominates (is practically equal to 1) and does not depend on $p$ (it may differ from 1 due to  relativistic corrections and the admixture of  exchange particles). On the contrary, in this article, we will discuss the systems in which the sum that includes the exchange particles $\sum_{n>2}N_n$ dominates. Since $\sum_n N_n=1$ and $N_n>0$, this means that  the probability $N_2$ of finding two constituents is small,  whereas the average number of  massless exchange particles can be large (maybe infinite). Referring to the content of the physical system, we  just mean the probability of finding a given number of particles in this system. 

Let us emphasize  that after the integration of  $|\psi_n(\vec{k}_1,\vec{k}_2,\ldots, \vec{k}_n;\vec{p})|^2$ over momenta  $\vec{k}_1,\vec{k}_2,\ldots, \vec{k}_n$, the dependence on $\vec{p}$ survives. That is, 
the probability $N_n(\vec{p})$ of finding $n$ particles depends on the total momentum $\vec{p}$ (it really depends on $|\vec{p}|$). It is different in different systems of reference.  This is related to the fact that the dependence of the relativistic wave function  $\psi_n(\vec{k}_1,\vec{k}_2,\ldots, \vec{k}_n;\vec{p})$, where $\vec{k}_1+\vec{k}_2+\ldots+\vec{k}_n=\vec{p}$, on the momenta 
is not reduced  (in contrast to the non-relativistic one) to its dependence on the relative momenta. That is, the center-of-mass motion is not separated, which is in contrast to the non-relativistic wave function. From a physical point of view, the reason is the fact that in a relativistic system, the interaction is not instantaneous; aside from the constituents, there is an indefinite number of  exchange quanta that are ``in flight''. Therefore, the constituent coordinates do not determine the position of the center of mass in the momentum space, which implies the impossibility of introducing the relative momenta. 

This can also be understood and confirmed  from a more formal point of view. As with a non-relativistic wave function, the relativistic state vector $|p\rangle$ satisfies the eigenstate equation 
\begin{equation}\label{Hp}
H|p\rangle=M|p\rangle,
\end{equation}
where $M$ is the total mass of the system. This equation implies that the system is in the rest frame, that is, the four-vector $p$ has the components 
$p=(M,\vec{0})$. Otherwise, the eigenvalue on  the r.h.-side of this equation contains the total energy $E_p=\sqrt{M^2+\vec{p}^2}$. The operator on the l.h.-side should be also determined; it is $\hat{P}^0$, the zero-component of the four-momentum operator $\hat{P}=(\hat{P}^0,\hat{\vec{P}})$. That is, in an arbitrary reference frame, this equation should be written in the form of four equations, with a separate equation for each component of the four-momentum operator $\hat{P}$:
$$
\hat{P}|p\rangle=p|p\rangle.
$$

The interaction $H_{int}$ enters into the operator $\hat{P}^0$ only; the operators $\hat{\vec{P}}$ are free.

As mentioned, in the decomposition (\ref{p}), each term corresponds to a fixed number of particles, whereas the interaction does not conserve the number of particles---they can be virtually created and annihilated.  In other words, the Hamiltonian $H$ (and $\hat{P}$) does not commute with the operator of the number of particles $\hat{N}$, which just results in  the superposition (\ref{p}) for its eigenvector $|p\rangle$. 
%The coefficients $\psi_n$ in this superposition depend on
%the eigenvalue $p$ of the operator $\hat{P}$.   This reflects the general property: in the decomposition $\phi_i=\sum_j C_{ij}\psi_j$ (here $\phi_i,\psi_j$ are %the eigenfunctions of non-commuting operators) the coefficients $C_{ij}$ depend on $i$.

Since the Fock components depend dynamically on $\vec{p}$, i.e., on the reference frame, one can try to find the most convenient reference frame. It turns out \cite{weinberg}  that this reference frame is the infinite momentum frame  $\vec{p}\to\infty$.  Its convenience is determined by the following reason. In general, the virtual particles are created not only by their emission from constituents (as a process of their exchange), but as a result of  vacuum fluctuations. A nucleon can virtually emit a meson: $nucleon\to nucleon + meson$. If so, the following virtual process is also possible: $vacuum \to nucleon+antinucleon+meson$ (or, e.g.,
$vacuum \to e^+e^-\gamma$). The advantage of the infinite momentum frame is in the fact that  vacuum fluctuations do not exist in this frame. When $\vec{p}\to\infty$, their energy tends to infinity.  Therefore, they are suppressed in the limit $\vec{p}\to\infty$ and do not contribute to the vacuum state vector. Hence,
in the infinite momentum frame,  the bare vacuum---an eigenstate of free Hamiltonian---is also an eigenstate of the Hamiltonian-containing interaction.
This leads to very considerable simplifications, not only of the vacuum state, but of the whole theory. 

The wave function in the infinite momentum frame is parametrized as follows. For brevity, we restrict ourselves to the two-body Fock component $\psi_2=\psi_2(\vec{k}_1,\vec{k}_2; \vec{p})$. When we go to the reference frame with $\vec{p}\to\infty$,
the transverse relative to the $\vec{p}\to\infty$ momenta $\vec{k}_{\perp 1,2}$ ($\vec{p}\cd\vec{k}_{\perp 1,2}=0$) does not vary with this transformation. We denote 
$\vec{k}_{\perp 1}\equiv\vec{k}_{\perp}$ ($\vec{k}_{\perp 2}=-\vec{k}_{\perp}$). Then, one introduces the ratios $x_{1,2}=\frac{k_{||1,2}}{p}$ (here, $\vec{k}_{||1,2} ||
\vec{p}$,\; $\vec{k}_{||1} +\vec{k}_{||2} = \vec{p}$)    and denotes $x_1\equiv x$, ($x_2=1-x$), $0\leq x\leq 1$. 
Therefore, in the limit $\vec{p}\to\infty$, the wave function in the infinite momentum frame is parametrized as $\psi_2=\psi_2(\vec{k}_{\perp},x)$.  Its contribution to the normalization integral reads:
\begin{equation}\label{norm2}
N_2=\frac{1}{(2\pi)^3} \int \left|\psi_2(\vec{k}_{\perp},x)\right|^2\frac{d^2k_{\perp} dx}{2x(1-x)}.
\end{equation}

Here, $N_2$  is the limiting value of $N_2(\vec{p}\to\infty)$. The $n$-body Fock component is similarly parametrized  (see, e.g., \cite{cdkm}).

These  features can be described  from a different, more formal and strict, but physically equivalent point of view.  
The Hamiltonian $H$ mentioned above determines the evolution of the state vector from one moment of time  to another.  In the 4D Minkowski space, this is evolution from one plane $t=const_1$ to another $t=const_2$. So far, we discussed the state vector defined on the equal-time plane $t=const$.
As is well known, two events that are simultaneous in one reference frame are not in equal time in other frame. Therefore, in addition to the plane $t=const$,  one can introduce a space-like plane of general orientation, say, defined by the equation $\alpha t-\beta z=const$, and, instead of transformation from one plane $t=const_1$ to another plane  $t=const_2$ (each in different moving reference frames), consider, now in the given reference frame,  the state vector defined on the planes of different orientations (but still the space-like ones). One such plane differs from the other by the parameters $\alpha,\beta$. The limiting case of the equal-time plane at $\vec{p}\to\infty$  for $\vec{p}||z$
corresponds to the light-front (LF) plane $t+z=const$ in this approach (it is enough  to take $t+z=0$). In this way, we come to the concept of  LF dynamics in which the state vector is defined on the LF plane $t+z=0$.

Another way to develop LF dynamics is to start with the LF Hamiltonian $\hat{P}^+=\hat{P}^0+\hat{P}^z$ instead of $\hat{P}^0$ and to consider the eigenstate equation in terms of this Hamiltonian. 
We obtain the same parametrization of the two-body Fock component $\psi_2=\psi_2(\vec{k}_{\perp},x)$ of the LF state vector.

If the state vector is defined on the equal-time plane $t=const$, we deal with the instant form of dynamics.  The approach in which the state vector is defined on the plane $t+z=const$ is called the LF dynamics. 
Another form of dynamics in which the state vector is defined on the hyperboloid  $t^2-\vec{x}^2=const$ is called the ``point form''.
All  three of these forms were introduced by Dirac \cite{dirac49}.
 The point form has also been successfully used in  physical applications \cite{klink2018}. 

Note, however, that the formulation of LF dynamics on the plane $t+z=0$ contains some inconvenience, since the 4D coordinates do not enter  this formulation  symmetrically. The coordinates $t,z$ are distinguished compared to $x,y$. This violates the explicit relativistic covariance. The explicitly covariant form was developed in \cite{VK76}. In this form, the LF hyperplane is determined by the equation $\omega\cd x =0$, where $\omega$ is a four-vector
$\omega=(\omega_0,\vec{\omega})$  such that $\omega^2=0$;  for a review, see \cite{cdkm}. In this version of LF dynamics, the explicit relativistic covariance is restored. The state vector still  depends on the orientation of the LF plane, as it should. Now it is reduced to the dependence on the four-vector $\omega$. The state vector  should now be written  as $|p, \omega \rangle$. {This considerably simplifies the calculations, especially when incorporating spins of particles, such as when finding  electromagnetic form factors.} % Please confirm that meaning has been retained
 In the particular case $\omega=(1,0,0,-1)$, we come back to the ordinary  version of LF dynamics.

Now, we can further {define what we mean  about the content of the system: We mean the  Fock components integrated over momenta and squared} % Please confirm that meaning has been retained 
$|\psi_n|^2$ for the state vector \emph{defined on the LF plane}. For the two-body contribution, this integral is given by Equation \mbox{(\ref{norm2}).}

%%%%%%%%%%%%%%%%%%%%%%%%%%%%%%
\section{Bethe--Salpeter Amplitude}\label{bsa}
Substituting the state vector $|p\rangle$ in the form of the decomposition (\ref{p}) into the LF eigenstate equation $\hat{P}^+|p\rangle=p^+|p\rangle$, one obtains the system of the integral equations for the Fock components $\psi_n$. A convenient method of deriving this system is to use the LF graph technique presented in \cite{cdkm}. In many known cases, the convergence of the decomposition~(\ref{p}) for the LF state vector is rather fast, and it can be truncated with good accuracy  \cite{LKMV_PLB,klsv2016}.   After truncation, the system of equations for the Fock components becomes finite and can be solved numerically. One should properly carry out the  renormalization. This approach has been developed in a series of articles;  for a review, see \cite{mstk}. 

However, if we expect that the system is dominated by a large number of exchange particles, this case is incompatible with fast convergence of the Fock decomposition.
Another approach to the theory of relativistic bound systems \cite{bs} deals not with the state vector $|p\rangle$ itself and its Fock decomposition, but with the matrix element taken from  the T-product of  the Heisenberg operators between the vacuum state and the state $|p\rangle$, namely:
\begin{equation} \label{bs}
\Phi(x_1,x_2,p)=\langle 0 \left| T\Bigl(\hat{\varphi}(x_1)\hat{\varphi}(x_2)\Bigr)\right|p\rangle\ ,
\end{equation}
where $\hat{\varphi}(x_{1,2})$ is the Heisenberg operator of the constituent field. The matrix element $\Phi(x_1,x_2,p)$ is the  BS  amplitude in the coordinate space. Sometimes, the amplitude defined by Equation (\ref{bs}) is called ``the two-body BS amplitude''. To avoid misunderstandings, we would like to emphasize that this is, to some degree, slang reflecting the fact that this BS amplitude depends on two variables. The state vector  $|p\rangle$ in the definition \mbox{(\ref{bs})} contains all of the Fock components, including the many-body ones. Therefore, the BS amplitude (\ref{bs}) implicitly incorporates  information not only about the two-body Fock sector, but also about the higher ones.

 The transformation
$$ 
\Phi(x_1,x_2,p)=(2\pi )^{-3/2}\exp \left[-ip\cd(x_1+x_2)/2\right]\tilde{\Phi}(x,p)\ ,\quad x=x_1-x_2\ ,
$$
\begin{equation}\label{bs3}
\Phi (k,p)=\int \tilde \Phi (x,p)\exp (ik\cd x)d^4x\ ,
\end{equation}
defines the BS amplitude  $\Phi (k,p)$ in the momentum space. It satisfies the BS equation:
\begin{equation}\label{bseq}
\Phi(k,p)=\frac{i^2}{\left[(\frac{p}{2}+k)^2-m^2+i\epsilon\right]\left[(\frac{p}{2}-k)^2-m^2+i\epsilon\right]}
\int \frac{d^4k'}{(2\pi)^4}iK(k,k',p)\Phi(k',p).
\end{equation}

For one-boson exchange in the spinless case, the kernel reads:
\begin{equation}\label{ladder}
iK(k,k',p)=\frac{i(-ig)^2}{(k-k')^2-\mu^2+i\epsilon}.
\end{equation} 

For  massless exchange, one should put in (\ref{ladder}) $\mu=0$.

%Solving the BS Equation (\ref{bseq}) for  $\Phi(k,p)$, one can find also the bound state mass $M^2=p^2$. 
It turns out that by knowing the BS amplitude $\Phi(k,p)$, one can extract from it the two-body Fock component $\psi_2$ corresponding to two constituents. 
This possibility is provided by the fact that the Heisenberg operators turn, on the quantization plane, into the Schr\"odinger ones, which are free and constructed from usual creation and annihilation operators:
\begin{equation} \label{bs5}
\hat{\varphi}(x)=\frac 1{(2\pi )^{3/2}}\int \left[a(\vec k)\exp (-ik\cd  x)+a^{\dagger}(\vec k)\exp (ik\cd  x)\right]\frac{d^3k}{\sqrt{2\varepsilon_k}}\ .
\end{equation}

This is true both for the equal-time case (when the Heisenberg operator $\hat{\varphi}(x)$ on the plane $t=0$ obtains the form (\ref{bs5})) and for the LF quantization (when  the Heisenberg operator $\hat{\varphi}(x)$ obtains the same form (\ref{bs5}) on the LF plane $\omega\cd x =0$). However, of course, this is not simultaneous: If the Heisenberg operator obtains the form (\ref{bs5}) on the LF plane, it has a very complicated form on the plane $t=0$ that is not reduced to (\ref{bs5}).  Therefore, if both arguments of the BS amplitude  $\Phi(x_1,x_2,p)$ are constrained by the LF plane $\omega\cd x_1=\omega\cd x_2=0$, then for the operators $\hat{\varphi}(x_{1,2})$, we can take Equation (\ref{bs5}), and the product of the two annihilation operators 
$a(\vec{k}_1)a(\vec{k}_2)$ contained in $\hat{\varphi}(x_1)\hat{\varphi}(x_2)$ is contracted with  the product of the two creation operators $a^{\dagger}(\vec{k'}_1)a^{\dagger}(\vec{k'}_2)$ contained in the two-body Fock sector of the state vector $|p\rangle$, and all of these operators disappear.  The result is proportional to the two-body Fock component $\psi_2$. The value of  $\Phi(x_1,x_2,p)$  in the coordinate space with the arguments constrained to the LF plane  $t_1+z_1=t_2+z_2=0$ corresponds, in the momentum space, to the integral from $\Phi(k,p)$ over $k_+$. In the explicitly covariant form, the relation between the BS amplitude  $\Phi(k,p)$ and the two-body Fock component obtains the form:
\begin{eqnarray}\label{lfwf}
\psi_2(\vec{k}_{\perp},x)=\frac{x(1-x)}{\pi\sqrt{N_{tot}}}\int_{-\infty}^{\infty} dy\, \Phi\left(k+\frac{y\omega}{\omega\cd p},p\right),
\end{eqnarray}
where $N_{tot}$ is the full normalization factor for a given state   that provides the normalization condition $F(0)=1$ of the elastic electromagnetic form factor for this state.  The derivation of the relation \mbox{(\ref{lfwf})} can be found in Ref. \cite{cdkm}, Section  3.3. The values of $N_{tot}$ in the limit of small binding energy for both normal and abnormal states are found analytically in Appendix \ref{app1}.

Solving the Equation (\ref{bseq})  for the BS amplitude $\Phi(k,p)$, one can find, with Equation \mbox{(\ref{lfwf})}, the two-body Fock component $\psi_2$,  and then, with Equation (\ref{norm2}), its contribution to the full norm of the state vector. Then, the contribution of the higher Fock sectors containing  the exchange particles (in addition to the constituents) is $N_{n> 2}=1-N_2$. Hence, in this way,  we can calculate the full contribution of the exchange particles, but not the contributions of the particular Fock sectors containing them.

%%%%%%%%%%%%%%%%%%%%%%%%%%%%%%%%%%
\section{Solving the BS Equation in the Wick--Cutkosky Model}\label{WCm}

For the massless ladder exchange, i.e., for the kernel (\ref{ladder}) with $\mu=0$, Wick and Cutkosky~\cite{wick,cutk}  reduced
the BS Equation (\ref{bseq}) to a one-dimensional equation that, in some limiting cases, can be solved analytically. The S-wave BS amplitude was represented in the~form
\begin{equation}\label{Phi}
\Phi_n(k,p)=\sum_{{\nu}=0}^{n-1}\int_{-1}^1g_{n}^{\nu}(z)dz  
\frac{-i m^{2(n-{\nu})+1}}{\left[m^2-\frac{1}{4}M^2 -k^2-p\cd k\,z-\imath\epsilon\right]^{2+n-{\nu}}}.       
\end{equation}

Here, $n$  is an integer parameter, and the solutions exist for any  $n=1,2,\ldots$
%In non-relativistic limit  this model reproduces to the well-known Coulomb bound state spectrum, in which $n$ is the principal quantum number. 
Substituting~(\ref{Phi}) into the BS Equation (\ref{bseq}) and following Ref. \cite{cutk}, one obtains the one-dimensional integral equation for $g_n^0(z)$:
\begin{equation} \label{gn}
g_n^0(z)=\frac{\alpha}{2\pi n}  \int_{-1}^1 {[R(z,z')]^n \over Q(z')}\; g_n^0  (z'),  
\end{equation}
where $\alpha$ plays the role of the eigenvalue and is related to the coupling constant $g$ in the kernel (\ref{ladder}) as follows\endnote{Do not confuse the coupling constant $g$ with the function $g_n^{\nu}(z)$ in the decomposition (\ref{Phi}) and with the solution $g_n^0(z)\equiv g_n^{\nu=0}(z)$  of Equation (\ref{gn}).}:
$$
\alpha=\pi \lambda = {g^2\over 16\pi m^2}
$$
and
\begin{equation}\label{Rzzp}
R(z,z')=\left\{
\begin{array}{ll}
\frac{1-z}{1-z'},& \mbox{for $z'<z$,}
\vspace{0.2cm}
\\
\frac{1+z}{1+z'},& \mbox{for $z'>z$,}
\end{array}
\right.
\end{equation}
\begin{equation}\label{Q}
 Q(z)=1-\eta^2(1-z^2),\quad  \eta^2=\frac{M^2}{4m^2}.
 \end{equation}
 
Namely, the single Equation (\ref{gn}) determines the mass spectrum. Other functions $g_n^{\nu}$ for the integer $0<\nu\leq n-1$
%, determining $\Phi(k,p)$ in (\ref{Phi}), 
satisfy the system of inhomogeneous integral equations in which the inhomogeneous term is determined by $g_n^0(z)$. These equations, as well as  Equation~(\ref{gn}), can be transformed into the differential form.   Together, along with Equation~(\ref{Phi}), they determine the BS amplitude $\Phi(k,p)$.
% whereas the eigenvalues, for given $n$, are determined by the single Equation (\ref{gn}).

The key point revealing the mathematical origin of the abnormal solutions is the following. Just as the homogeneous Schr\"odinger equation can have a few (or even infinite) eigenvalues, the homogeneous Equation (\ref{gn}), for a given value of $n$ and the coupling constant $\alpha>\frac{\pi}{4}$, also does not have a single solution, but an infinite set of eigenfunctions and corresponding binding energies \cite{wick,cutk}.
%It tuned out \cite{wick,cutk} that for fixed $n$, the Equation (\ref{gn}), corresponding to the massless exchange kernel (\ref{ladder}) 
%has an infinite number of solutions,
To label them, an extra quantum number \mbox{$\kappa=0,1,2,\ldots$} is introduced.
%, which also labels the corresponding discrete spectrum of mass squared eigenvalues $M_{n\kappa}^{2}$. 
%We use the notation $g^{0}_{n\kappa}$ to identify a particular solution with the mass  $M_{n\kappa}^{2}$.
The solution {$g_{n\kappa}^0(z)$} with the mass  $M_{n\kappa}^{2}$ vs. $z$ has $\kappa$ nodes within the interval
$-1<z<1$  and a definite parity  \cite{cutk}:
$$
g^0_{n\kappa} (-z)= (-1)^{\kappa} g^0_{n\kappa}(z).
$$
%The parity  is also preserved inside the  ensemble  $\{ g_{n\kappa}^{\nu} \}$
%when varying $\nu=0,1,\ldots$  and this entails, through Eq. (\ref{Phi}),
%that for even (odd) values of $\kappa$,

Hence, the BS amplitude $\Phi(k,p)$ defined by Equation (\ref{Phi}), which is taken in the rest frame as a function of the relative
energy $k_0^{}$, is  even or odd.

The odd solutions  do not contribute to  the $S$-matrix  \cite{cia,nai}.
Therefore, we will mainly concentrate on the solutions with even $\kappa$, which
may have a physical meaning.
 
 A particular set of solutions with $\kappa=0$ and arbitrary $n$, with a small binding energy $B_n \ll m$, reproduces the non-relativistic Coulomb 
spectrum \cite{wick,cutk} in the potential \mbox{$V(r)=-\frac{\alpha}{r}$}:
%for the function $g_{n0}^{0}(z)$, 
%determining the BS amplitude (\ref{Phi}), 
%Wick and Cutkosky \cite{wick,cutk} reproduced, for $\kappa=0$, the Coulomb spectrum, i.e. the Balmer series for the Schr\"odinger equation with the %potential $V(r)=-\frac{\alpha}{r}$:
\begin{equation}\label{Balmer}
B_n=\frac{m\alpha^2}{4n^2}.
 \end{equation}
 
For the ground-state solution $n=1$, for $B_1\ll m$, Wick and Cutkosky \cite{wick,cutk} found $g_{10}^{0}(z)=1-|z|$. The corresponding 
two-body Fock component $\psi_2$ is expressed through the solution $g(z)$ below in  Section \ref{2b}. It is given by Equation (\ref{wf1}). In the non-relativistic limit, it obtains the form (\ref{wf1a}) and coincides with the non-relativistic ground-state Coulomb wave function.

On the other hand, the solutions with non-zero $\kappa=1,2,\ldots$ are completely decoupled from the non-relativistic solutions.
They have a true relativistic nature and have no non-relativistic counterparts. These solutions are called  ``abnormal''.

\begin{figure}[H]
%\vspace{0.8cm}
\includegraphics[width=10.5 cm]{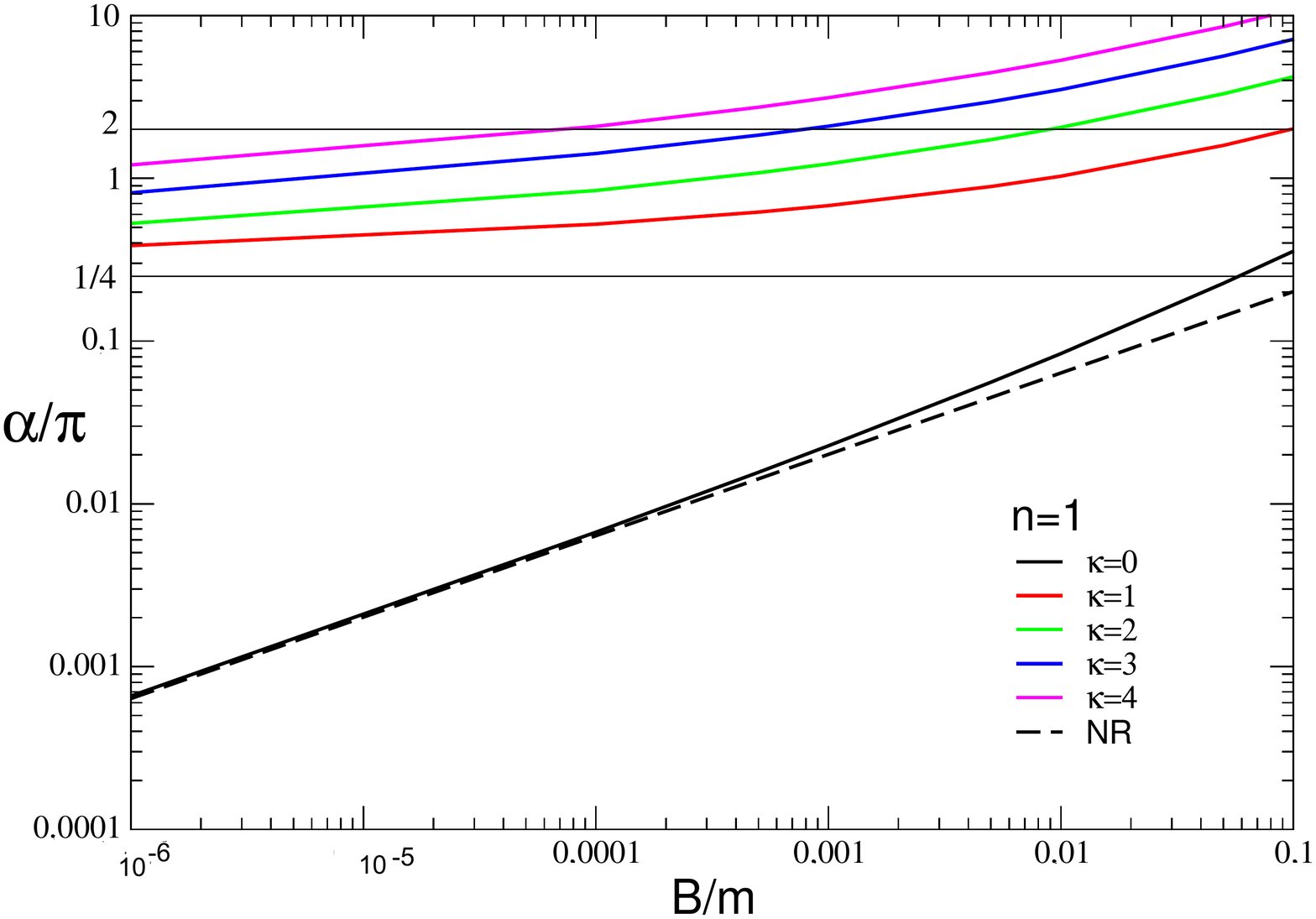}
\caption{{Spectrum in the Wick--Cutkosky} %MDPI: 1. Please use scientific notation. (e.g., 8 × 10³, not 8E3); 2. please change the comma in digits to dot (e.g., 0,1 should be 0.1)
 model  of the lowest coupling constants $\alpha(B)/ \pi$ vs. $\kappa$ for $n=1$.
The black solid line corresponds to $\kappa=0$. The black dashed line is the non-relativistic solution. The colored solid lines correspond to  $\kappa=1\div 4$. The horizontal line at  $\alpha/\pi=1/4$  corresponds   to the minimal coupling constant   for which  the abnormal solutions exist. ({Adapted from} %MDPI: Please make sure that permission has been obtained and there is no copyright issue. 
 \protect{\cite{cks21}).}\label{lambda_B}}
\end{figure}

In  Figure \ref{lambda_B}, which was taken from Ref. \cite{cks21}, we show, in the Wick--Cutkosky model, the behavior of the coupling constant vs. the binding energy for a few of the lowest states with $n=1$. The black solid line corresponds to $\kappa=0$, whereas the black dashed line is the non-relativistic solution. With the increase in $B$,
they deviate from each other due to the relativistic correction, which is logarithmic. The colored solid lines correspond to  $\kappa>0$. The non-relativistic solutions that could be associated with them do not exist. 

For the small binding energy $B\to 0$, the following approximate formula for the spectrum of abnormal states with $\kappa=2,3,\ldots$ was found \cite{wick,cutk}:  
\begin{equation} \label{abnspectr}
  M_{n\kappa}^2\simeq 4m^2\left[   1 - \exp\left( -\frac{(\kappa-1)\pi}
   {\sqrt{{\alpha\over \pi}-{1\over 4}}}\right)\right].    
\end{equation}

It is evident that for real eigenvalues, the condition $\alpha>\pi/4$  must be satisfied. In this approximation ($B/m\ll 1$),  the bound-state mass $M_{n\kappa}^2$ depends on $\kappa$ only, but  does not depend on $n$. If $\alpha\to\pi/4$, all of the abnormal excited energies tend to 0. In order for them to be distinguishable from the continuum, the coupling constant $\alpha$ should not be too close to $\pi/4$, but should be at least $\alpha\approx 4\div 5$.

The  binding  energies for two normal ($n=1,2$, $\kappa=0$) and four abnormal  ($n=1,2$, $\kappa=2,4$) states
found through the numerical solution of Equation (\ref{gn})  are shown in Table \ref{tab1}.

\begin{table}[H]
\small
\caption{Binding energy ${\bf B}$ (in units of {m}%MDPI: unit should not be italics, changed all to normal, please confirm
) for the quantum numbers $n=1,2$ and $\kappa=0,2,4$. ${\bf N_2}$ is the contribution of these states to the full norm. The calculations were carried out for the coupling constant $\alpha=5$. ({Adapted from  Tables 1 and 2 from} %MDPI: Please make sure that permission has been obtained and there is no copyright issue. 
 \cite{cks21}). \label{tab1}}
\setlength{\tabcolsep}{9.2mm}
\begin{tabular}{cccll}
\toprule
\textbf{No.}&   $\textbf{{n} %MDPI: please confirm if n should be italics
}$ & {$\bm\kappa$} &  ${\bf B}$ & ${\bf N_2}$ \\
%\textbf{Title 1}	& \textbf{Title 2}	& \textbf{Title 3}\\
\midrule
1&1&0  &   0.999259         &0.65 \\
%Entry 1		& Data			& Data\\
2&2&0  &   0.208410         & 0.61 \\
%Entry 2		& Data			& Data\\
%%%%%%%%%%%%%%%%%%%%%%%%%
3& 1 &2  &
$3.51169 \times 10^{-3}$
& 0.094 \\
4& 2 &2 &
$1.12118 \times 10^{-3}$
& 0.077 \\
5&1  &4&
$1.54091\times 10^{-5}$ & $6.19\times 10^{-3}$ \\
%Entry 1		& Data			& Data\\
6&2  &4&
$4.95065    \times 10^{-6}$ &
$2.06\times 10^{-5}$\\
\bottomrule
\end{tabular}
\end{table}

One can see that even for $\alpha=5$, the binding energies of  abnormal states {Nos.} % Should this be a dash rather than a \div sign?
 $3\div 6$ are rather small:  $B$$\sim$$10^{-3}\div 10^{-6}$ m 
relative to the normal ones. The normal states can have such small binding energies if they are extremely excited.  Column ${\bf N_2}$, which shows the two-body contributions---the main subject of this review, together with the many-body one---will be discussed below.

\begin{figure}[H]
%\vspace{0.8cm}
\includegraphics[width=10.5 cm]{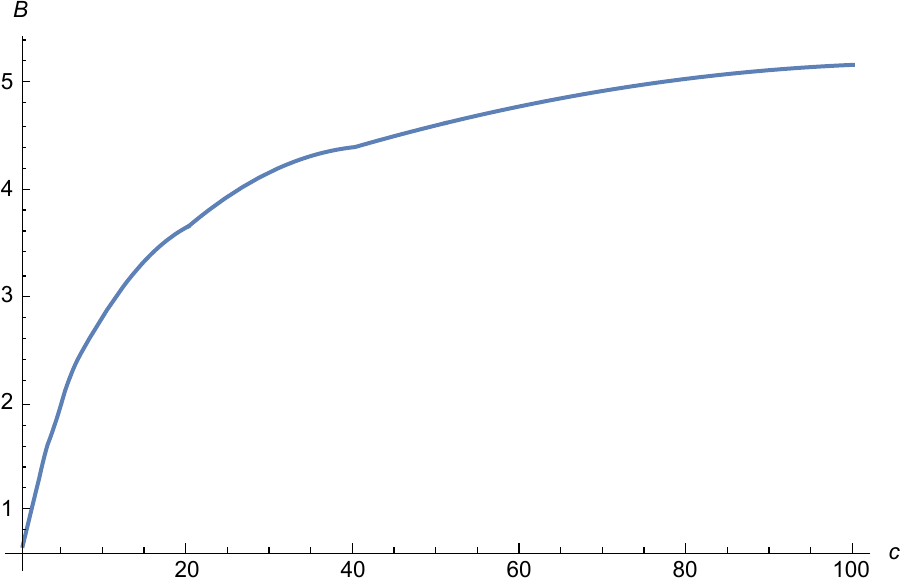}
\caption{Dependence of the binding energy $B$ of the state $n=1$, $\kappa=0$ (normal) for $\alpha=5$ on the speed of light~$c$.} \label{Bcn}
\end{figure}

\begin{figure}[H]
%\vspace{0.8cm}
\includegraphics[width=10.5 cm]{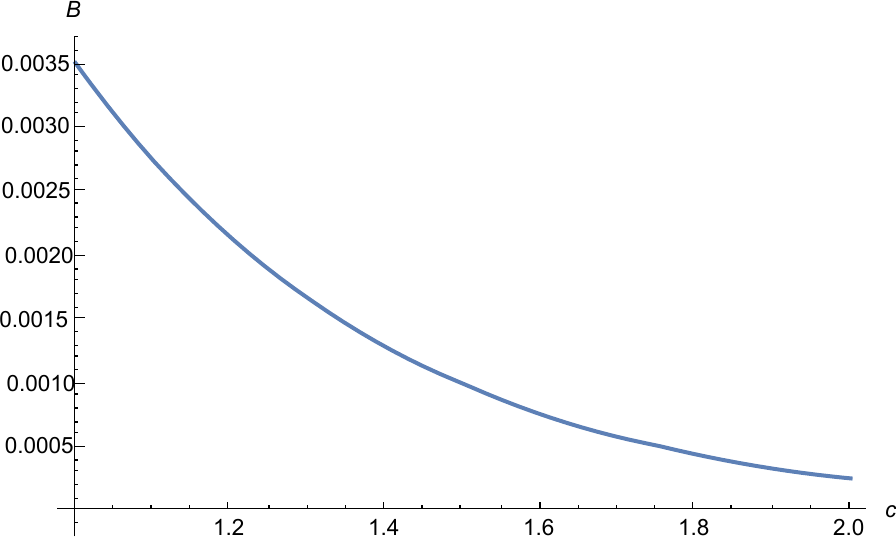}
\caption{Dependence of the binding energy $B$ of the abnormal state $n=1$, $\kappa=2$ for $\alpha=5$ on the speed of light~$c$. \label{Bca}}
\end{figure}

The normal and abnormal solutions drastically differ in their behavior in the non-relativistic limit. The binding energy of normal solutions tends to a finite limit, whereas the abnormal solutions disappear---as mentioned, they have no non-relativistic counterparts. The latter means that in the non-relativistic limit, they are pushed out of the discrete spectrum. The non-relativistic limit means that all of the velocities are much smaller than the speed of light $c$, which, in the true non-relativistic realm, is considered as infinite. Therefore, it is convenient to find the non-relativistic limit  by recovering the speed of light $c$ (which was put to 1) in  Equation (\ref{gn}) while considering $c$ as a parameter and taking the limit $c\to \infty$. For this aim, 
we should introduce $c$ in the input parameters, i.e., replace  $m\to mc^2$, $\alpha=\frac{e^2}{\hbar c}\to \frac{\alpha}{c}$. We \emph{should not} replace $M\to Mc^2$, since $M$ is the output calculated with the $c$-dependent input parameters. The dependence of $M$  (as well as the dependence of $B$) on $c$  is determined by the equation. The dependence of the binding energy $B$ on the speed of light $c$ for the normal state $n=1$, $\kappa=0$ ($\alpha=5$, No. 1 in  Table \ref{tab1} for $c=1$) is given in Figure \ref{Bcn}. It shows  what happens with the binding energy in a smooth transition from the relativistic approach to the non-relativistic one.
 %At $c\to \infty$ it tends to the nonrelativistic value.
This dependence is shown in Figure \ref{Bca} for the case of the abnormal state $n=1$, $\kappa=2$ ($\alpha=5$, No. 3 in  Table \ref{tab1} for $c=1$).
At $c\to\infty$, in Figure \ref{Bcn}, the binding energy of the normal state of the BS equation tends to be constant (which %for $\alpha\ll 1$ 
is given by the Schr\"odinger equation, i.e.,  by  the Balmer series in Equation (\ref{Balmer})). However, Figure \ref{Bca} shows that  the binding energy of the abnormal state has a quite different behavior: It decreases and tends to zero when $c$ increases. That is, the abnormal state disappears in the non-relativistic limit.

\begin{figure}[H]
%\vspace{0.8cm}
\includegraphics[width=10.5 cm]{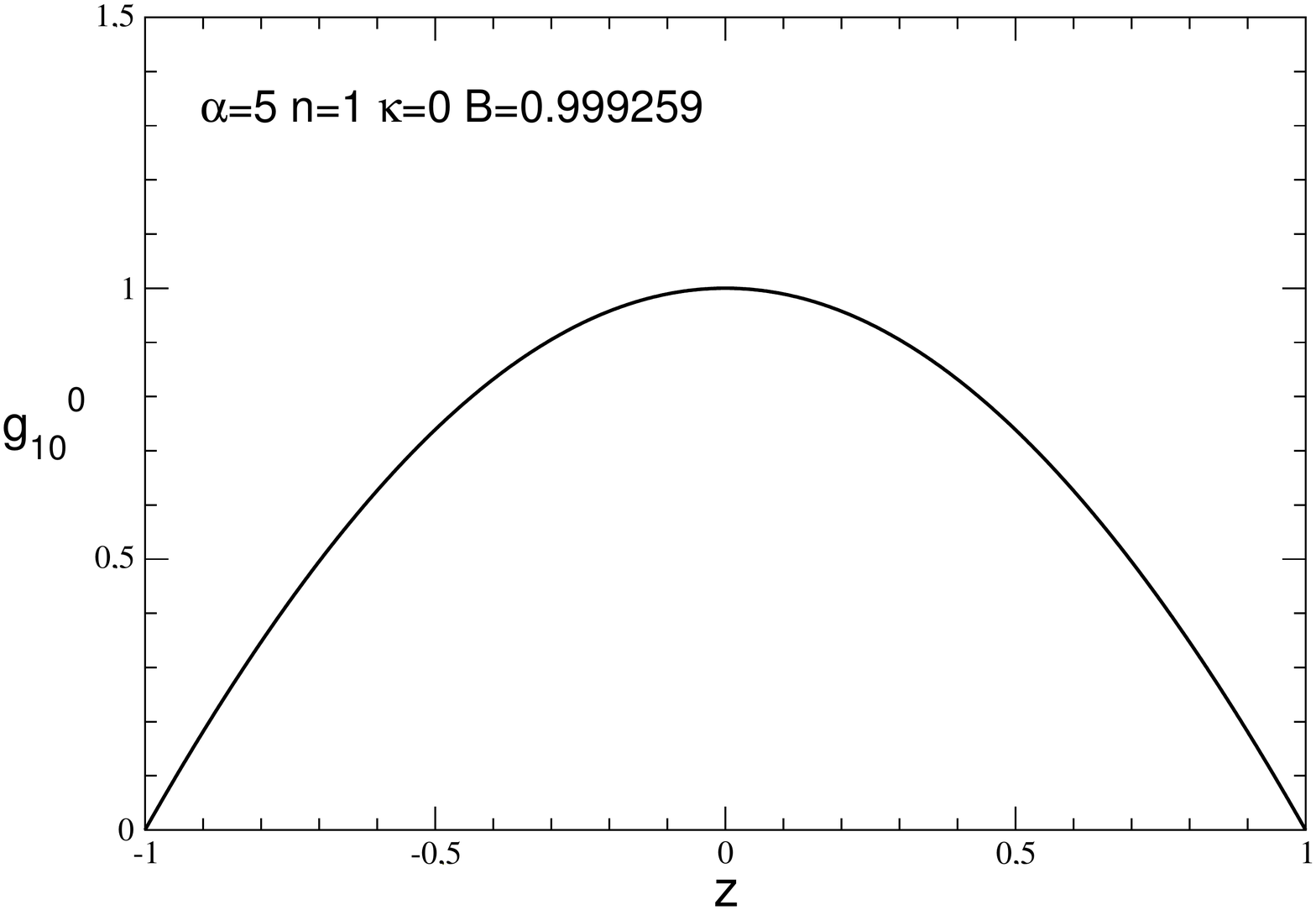}
\caption{{$g_{10}^0$ for} %MDPI: 1. please change hyphen (-) to minus sign; 2. please change the comma in digits to dot (e.g., 0,1 should be 0.1)
   state No. 1 ($\kappa=0$, normal) from  Table \ref{tab1}. ({Adapted from} %MDPI: Please make sure that permission has been obtained and there is no copyright issue. 
 {\cite{cks21}).}\label{fig1}}
\end{figure}

\begin{figure}[H]
%\vspace{0.8cm}
\includegraphics[width=10.5 cm]{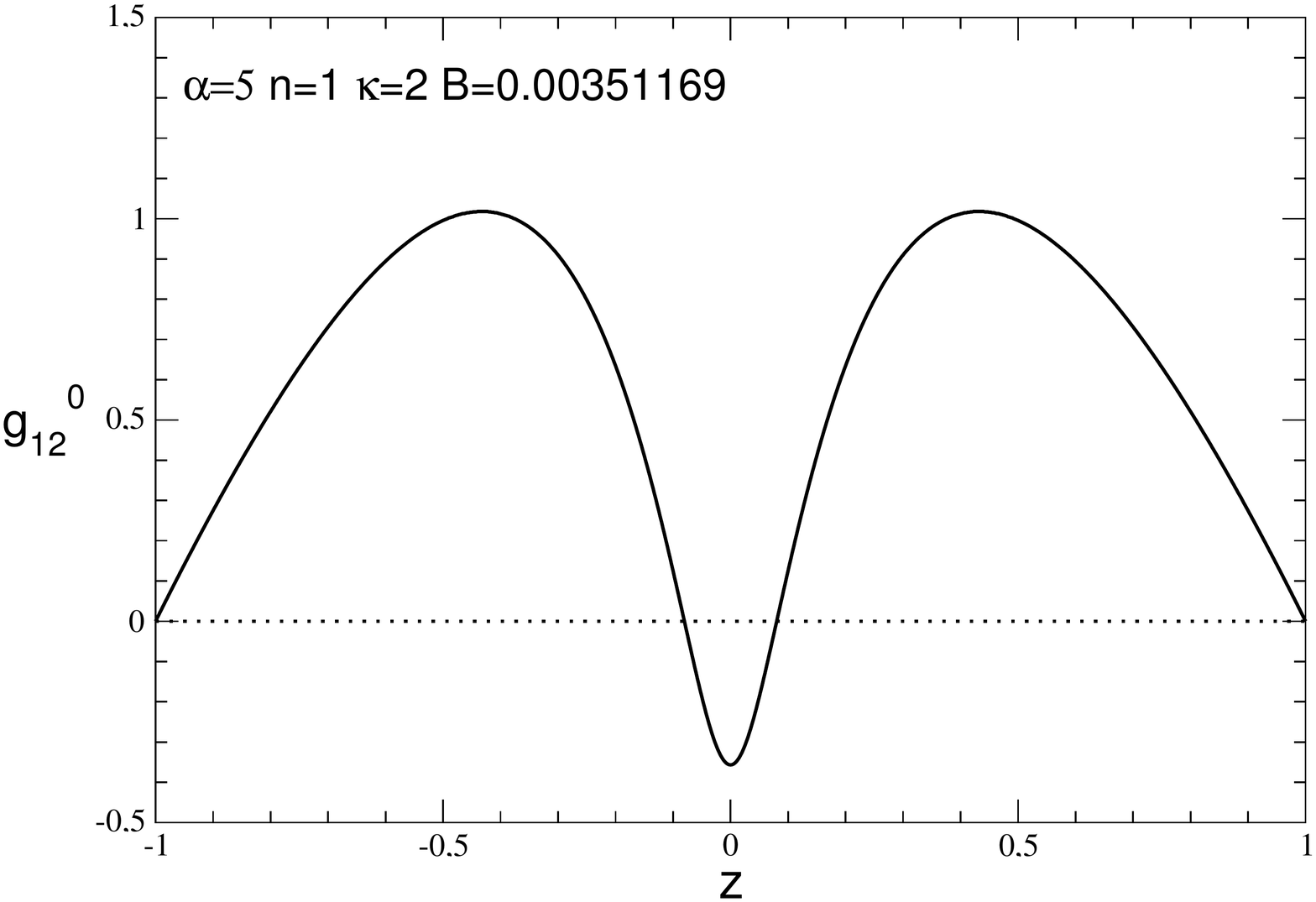}
\caption{{$g_{12}^0$ for state} %MDPI: 1. please change hyphen (-) to minus sign; 2. please change the comma in digits to dot (e.g., 0,1 should be 0.1)
 No. 3 ($\kappa=2$, abnormal)  from Table \ref{tab1}. ({Adapted from} %MDPI: Please make sure that permission has been obtained and there is no copyright issue. 
 \protect{\cite{cks21}).} \label{fig1p}}
\end{figure}

\begin{figure}[H]
%\vspace{0.8cm}
\includegraphics[width=10.5 cm]{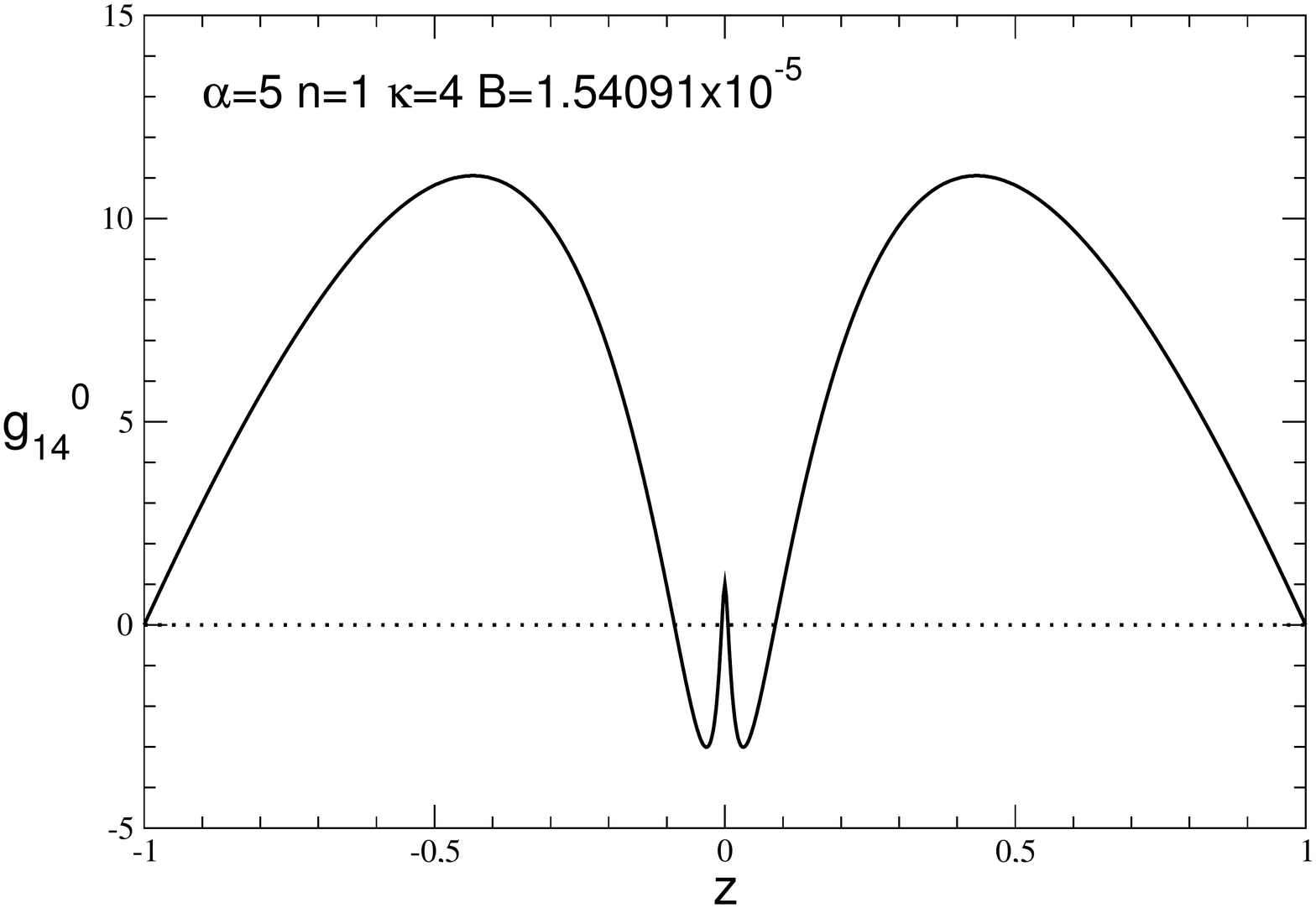}
\caption{{$g_{14}^0$ for state} %MDPI: please change hyphen (-) to minus sign. 
 No. 5 ($\kappa=4$, abnormal)  from Table \ref{tab1}. ({Adapted from} %MDPI: Please make sure that permission has been obtained and there is no copyright issue. 
 \protect{\cite{cks21}).}
%{\bf Faute! $\kappa=2\to\kappa=4$!} 
\label{fig10_14}}
\end{figure}

The solutions for the $n=1$ and $\kappa=0,2,4$ states---$g_{10}^0$,
$g_{12}^0$, and $g_{14}^0$, arbitrarily normalized---are displayed in  Figures \ref{fig1}--\ref{fig10_14}.
 \vspace{-9pt}

In  Figures \ref{fig1}--\ref{fig10_14}, we see that the number of nodes of each function $g(z)$ indeed coincides with the value of $\kappa$.

%We represent now the equation determining the function $g^{1}_{2\kappa}$, i.e.,  for $n=2$ and $\nu=1$. 
As mentioned, the functions $g^{0}_{n\kappa}(z)$ for  $\nu=0$ and for any $n$ satisfy the Equation~{(\ref{gn})}, which determines a series of eigenfunctions 
and binding energies labeled by the quantum number $\kappa$. For $n=2$, the BS amplitude (\ref{Phi}) 
contains another function $g^{1}_{2\kappa}$, corresponding to $\nu=1$. The equation determining this function can be also found through the substitution of (\ref{Phi}) into the BS equation \mbox{(\ref{bseq})}. It has the form (see, e.g., \cite{cks21}): 
\begin{equation}\label{g21}
g_2^1(z)=   \frac{\alpha}{6\pi}\int_{-1}^1\frac{R(z,z') } {[Q(z')]^2}g_2^{0}(z')dz'    
+   \frac{\alpha}{2\pi}\int_{-1}^1\frac{R(z,z')}{Q(z')}  \, g_2^{1}(z')dz'.
\end{equation}

$R(z,z')$ and $Q(z')$ are defined in Equations (\ref{Rzzp}) and (\ref{Q}). They are inhomogeneous relative to $g_2^1(z)$; the inhomogeneous term is determined 
by the function $g_2^{0}$, which is known from Equation \mbox{(\ref{gn})}.

 Then, the BS amplitudes for $n=1,2$ are expressed via these solutions, according to Equation (\ref{Phi}),  as:
\begingroup
\makeatletter\def\f@size{9}\check@mathfonts
\def\maketag@@@#1{\hbox{\m@th\normalsize\normalfont#1}}%
\begin{eqnarray}\label{Phi1}
\Phi_1(k,p)&=&\int_{-1}^1 \frac{-im^{3}g_{1}^{0}(z)dz}{\left[m^2-\frac{1}{4}M^2 -k^2-p\cd k\,z-\imath\epsilon\right]^{3}},       
%\end{equation}
%\begin{equation}
\\
  \Phi_2(k,p)&=&
   \int_{-1}^1\frac{-im^3 \,g_{2}^1(z)dz}
      {\left[m^2-\frac{1}{4}M^2 -k^2-p\cd k\,z-\imath\epsilon\right]^{3}}
  +
  \int_{-1}^1\frac{-i m^5\,g_{2}^0(z)dz}
      {\left[m^2-\frac{1}{4}M^2 -k^2-p\cd k\,z-\imath\epsilon\right]^{4}}.
  \label{Phi2}
\end{eqnarray}\endgroup

The system of equations for  $g_{n\kappa}^{\nu}(z)$ with  any $n,\nu$ is given in \cite{cks21}.

%%%%%%%%%%%%%%%%%%%%%%%%%%%%%%%%%%%%%%%%%%%%%%
\section{Two-Body Contributions}\label{2b}
Knowing the BS amplitude $\Phi(k,p)$, we can find  the two-body Fock component $\psi(\vec{k}_{\perp},x)$ with  Equation (\ref{lfwf}), and then, with Equation (\ref{norm2}), we can find 
its contribution $N_2$ to the normalization of the full state vector. The difference $1-N_2$ determines the contribution of the many-body Fock sectors with $n>2$. With this method, the content of the normal states up to an extremely relativistic binding energy was analyzed in Ref. \cite{hvk2004}. In \cite{cks21}, we applied this method to the abnormal states.

We substitute Equations (\ref{Phi1}) and (\ref{Phi2}) into Equation (\ref{lfwf}) and integrate over $\beta$. We omit the technical details of this integration, which can be found in Ref. \cite{cks21}, Appendix A. For $n=1$ (ground state), the result reads:
\begin{equation}\label{wf1}
\psi_{n=1}(\vec{k}_{\perp},x)=\frac{m^3x(1-x)g_1^0(1-2x)}{\sqrt{N_{tot}}[\vec{k}_{\perp}^2+m^2-x(1-x)M^2]^2}
\end{equation}

We will show that in the non-relativistic limit, in appropriate variables, this wave function reproduces the ground-state hydrogen wave function.
Instead of the pair of variables $k_{\perp},x$, we introduce another pair $q,\theta$ with the formulas:
$$
k_{\perp}=q\sin\theta,\; x=\frac{1}{2}\left(1-\frac{q\cos\theta}{\sqrt{m^2+q^2}}\right).
$$

Then, for $q\ll m$ and  for the integration volume in (\ref{norm2}), we obtain:
$$
\frac{d^2k_{\perp} dx}{2x(1-x)}=\frac{2\pi k_{\perp}dk_{\perp} dx}{2x(1-x)}=
\frac{2\pi q^2dq\sin\theta d\theta}{\sqrt{m^2+q^2}}\approx \frac{2\pi q^2dq\sin\theta d\theta}{m}. 
$$

We move the factor $m$ from this denominator  to the wave function, multiplying the latter by $1/ \sqrt{m}$. 
 Instead of  $M$, we also introduce the binding energy $B$: $M=2m-B$. As already indicated above, for $B \to 0$, $g(z)=1-|z|$, which gives $g_1^0(1-2x)|_{x\approx 1/2}\approx 1$. At last,  for $N_{tot}$, we take expression (\ref{Ntot2}) from Appendix \ref{app1}. Then, the wave function (\ref{wf1})  takes the form:
 \begin{equation}
\psi_{n=1}(q)=\frac{8\sqrt{\pi}(Bm)^{5/4}}{(q^2+Bm)^2}.
\label{wf1a}\\
\end{equation}

The wave function $\psi_{n=1}(q)$ (Equation (\ref{wf1a})) is just the ground-state hydrogen wave function in the momentum space.

 The normalization condition (\ref{norm2}) turns into:
 \begin{equation}
 N_2=\frac{1}{(2\pi)^3}\int \psi^2_{n=1}(q)2\pi q^2dq\sin\theta d\theta=1.
 \label{norm2a}
\end{equation}

Let us emphasize that the normalization condition (\ref{norm2a}) is not imposed, but derived. What is imposed (via condition $F(0)=1$ applied to the electromagnetic form factor; see Section \ref{ffs}) is the normalization condition (\ref{normstv}) for the full state vector. However, Equation~(\ref{norm2a}) is a consequence of calculating the two-body wave function with \mbox{Equations~(\ref{lfwf})} and (\ref{wf1}) and calculating $N_{tot}$ with Equations~\mbox{(\ref{Ntot1})} and  \mbox{(\ref{Ntot2})}. Its coincidence with 1 shows that, in the non-relativistic limit, {the system consists of two constituents at 100\%, as expected.} As we will see below, it is not so for the relativistic~states. % Please confirm that meaning has been retained

Similarly, for the first excited state $n=2$,
\begin{equation}\label{wf2}
\psi_{n=2}(\vec{k}_{\perp},x)=\frac{m^3x(1-x)g_2^1(1-2x)}{\sqrt{N_{tot}}[\vec{k}_{\perp}^2+m^2-x(1-x)M^2]^2}
+\frac{2m^5x(1-x)g_2^0(1-2x)}{3\sqrt{N_{tot}}
[\vec{k}_{\perp}^2+m^2-x(1-x)M^2]^3}.
\end{equation}

This wave function can be also transformed into the form of the solution of the Schr\"odinger equation with the Coulomb potential  for the excited  $n=2$ state. 
%Like the ground state wave function (\ref{wf1}) is transformed,
Substituting the wave functions (\ref{wf1}) and (\ref{wf2}) into (\ref{norm2}), we obtain the contributions of the two-body sectors for these states \cite{cks21}:
\begin{equation}\label{norm10}
N^{n=1}_2=\frac{1}{384 \pi^2 N_{tot}}\int_{-1}^1\frac{(1-z^2)[g_1^0(z)]^2dz}{[Q(z)]^3},
\end{equation}
\begin{equation}\label{norm21}
N^{n=2}_2=\frac{1}{3\cdot 2^7\pi^2  N_{tot}}\int_{-1}^1dz\,(1-z^2)
\left\{\frac{[g_2^1(z)]^2}{[Q(z)]^3}\right.
+\left.\frac{g_2^1(z)g_2^0(z)}{[Q(z)]^4}
+\frac{4}{15}\frac{[g_2^0(z)]^2}{[Q(z)]^5}\right\}
\end{equation}
with $Q(z)$ being defined in Equation (\ref{Q}). 
%The value $N_{tot}$ is found from condition of normalization of the elastic electromagnetic form factor
%$F(0)=1$. It is calculated analytically in the Appendix \ref{***} in the limit of small binding energy, both for normal and abnormal states.  

\begin{figure}[H]
%\vspace{0.8cm}
\includegraphics[width=10.5 cm]{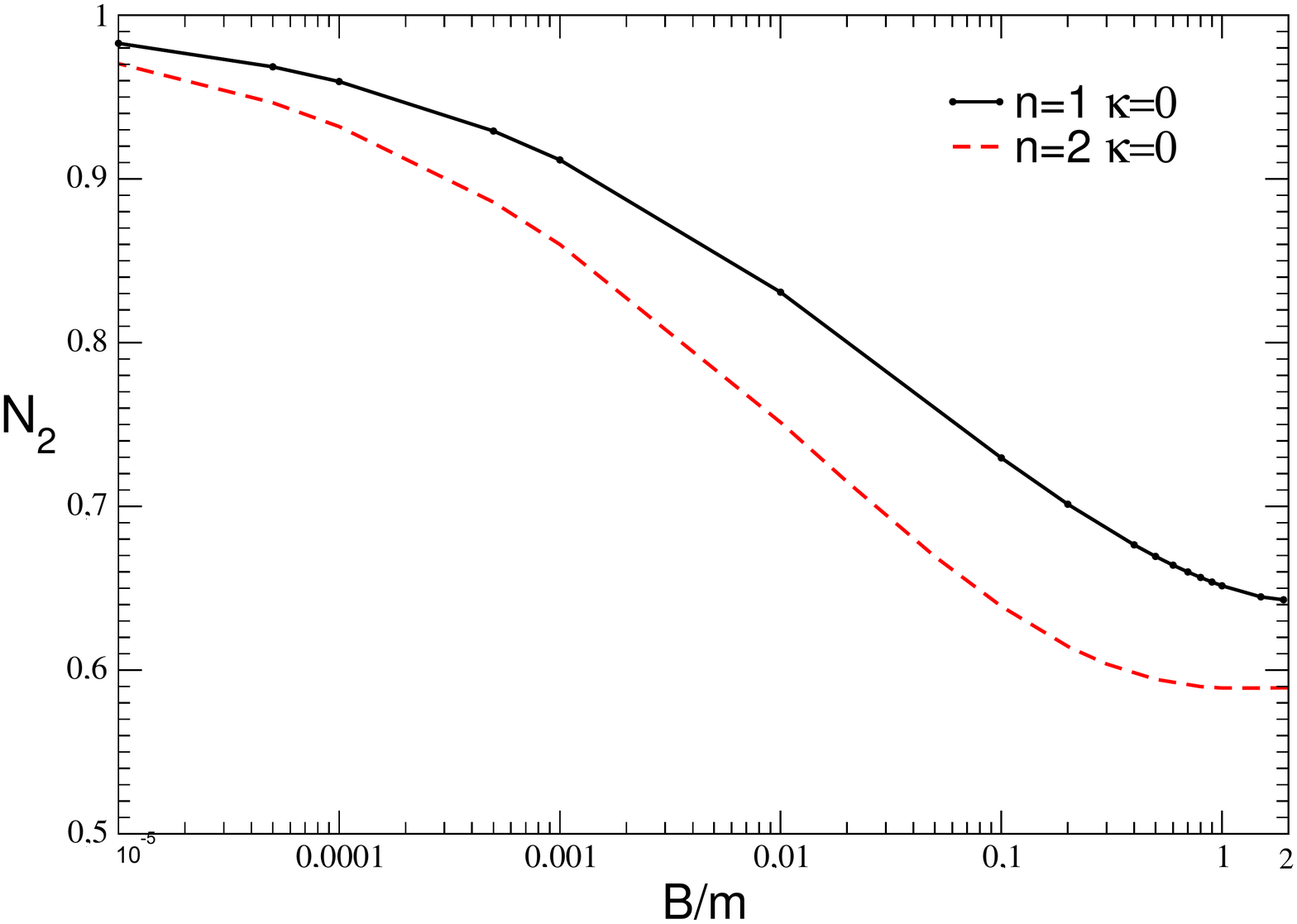}
\caption{{The two-body} %MDPI: 1. Please use scientific notation. (e.g., 8 × 10³, not 8E3); 2. please change the comma in digits to dot (e.g., 0,1 should be 0.1)
 contribution $N_2$  vs. the binding energy $B$ for
  the  normal ($\kappa=0$, $n=1$, and $n=2$) states. ({Adapted from} %MDPI: Please make sure that permission has been obtained and there is no copyright issue. 
 \protect{\cite{cks21}).} \label{N2Bnorm}}
\end{figure}

The numerical results are given in the last column of  Table \ref{tab1}. For the normal states, Nos.1 and 2, the two-body contribution  for
$\alpha=5$ is around 60\%. This value, which considerably deviates from 100\%, is related to the rather large binding energy ($B$$\sim$$0.2\div 1$ m) and, hence, to the considerable relativistic effects.  The dependence of $N_2$ for the normal states $n=1,2$, $\kappa=0$,   on the binding energy $B$ is presented in Figure \ref{N2Bnorm}.
%\noindent

For the abnormal state $n=1, \kappa=2$,
as the calculations show, the value of $N_2$ increases with the increase in $\alpha$. One can ask: Is it possible for some adequately large $\alpha$ to obtain an abnormal state that is not dominated by  exchange particles, but  with   significant two-body content $N_2$, say, 50\%? 
It turns out that this is impossible: The abnormal states are always dominated by the exchange particles for any physically admissible values of the coupling constant
$\alpha$ and the corresponding binding energies. With the increase in $\alpha$, the squared ground- (normal-) state mass  $M^2$ quickly decreases, and at $\alpha=2\pi$, it reaches the value $M^2=0$. For larger $\alpha$, $M^2$ becomes negative, and the system cannot be considered as the physical one. Though $M^2$ remains positive  for the abnormal (excited) states, for such large values of $\alpha$, these states (though with positive $M^2$) are the excited states of the physically senseless ground state with $M^2<0$, and therefore, in our opinion, they also have no  physical meaning. For example, for the limiting value $\alpha=2\pi$ ($M_{ground}=0$),  we find  $B=0.00903$ m and $N_2=0.156$ for the first abnormal state. This is the maximal value of $N_2$ that can be achieved for the abnormal state. If we continue to increase $\alpha$, then for 
$\alpha=11$, we obtain $B=0.059$ m and $N_2=0.55$~m$^2$ for the abnormal state, i.e., an approximately 50--50\% relation between the contributions of the constituent and exchange particles. However, the squared ground-state mass  becomes $M^2=-3.94$ m$^2$, so the system loses all physical~meaning.

When  $\alpha\to 0$, the constituent contribution for the ground state reads \cite{hvk2004}:
$$
N_2=1-\frac{2\alpha}{\pi}\log\frac{1}{\alpha}.
$$

When it is rewritten in terms of the binding energy, $B=\frac{1}{4}\alpha^2m$, $N_2$ obtains the form:
\begin{equation}\label{N2a}
N_2(B\to 0)=1+\frac{1}{\pi}\sqrt{\frac{4B}{m}}\log\frac{4B}{m}
%\stackrel{B\to 0}{\longrightarrow} 1.
\end{equation}

\begin{figure}[H]
%\vspace{0.8cm}
\includegraphics[width=10.5 cm]{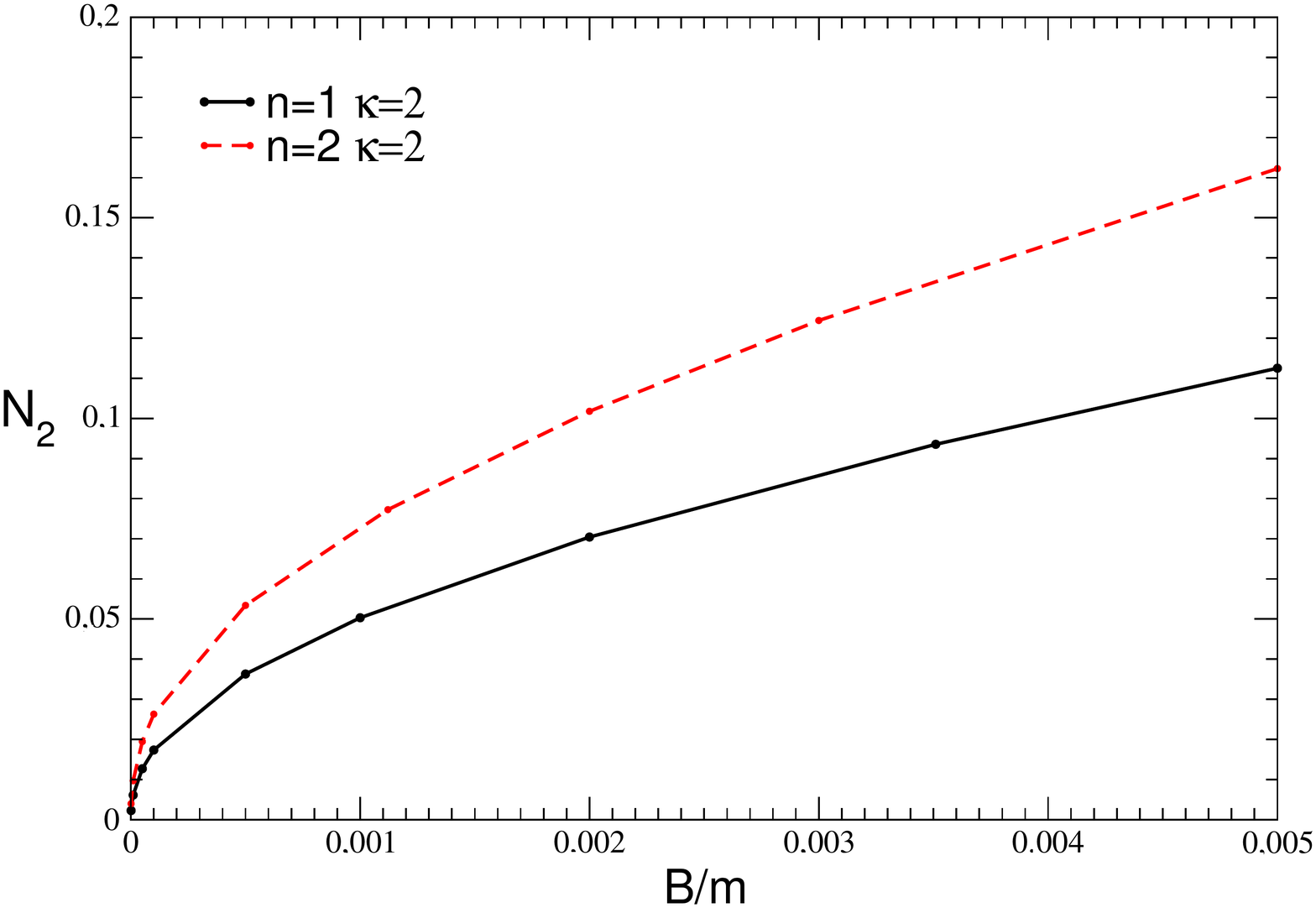}
\caption{{The two-body} %MDPI: please change the comma in digits to dot (e.g., 0,1 should be 0.1)
 contribution $N_2$  vs. the binding energy $B$ for
  the  abnormal ($\kappa=2$, $n=1$, and $n=2$) states. ({Adapted from} %MDPI: Please make sure that permission has been obtained and there is no copyright issue. 
 \protect{\cite{cks21}).} \label{N2Babnorm}}
\end{figure}

On the contrary, as seen in Table \ref{tab1}, for the abnormal states, the two-body contribution is rather small: $N_2\approx 10^{-1}\div 10^{-5}$ (for $n=1, \kappa=2$ and $n=2, \kappa=4$, respectively).
The dependence of $N_2$ for the abnormal states $n=1,2$, $\kappa=2$,  on the binding energy $B$ is shown in Figure \ref{N2Babnorm}.
For $B\to 0$, one can derive the analytical formula for $N_2$ for the abnormal states. For the particular case of $n=1,\kappa=2$, the derivation is given in Appendix \ref{app1}. The dominating term reads:
\begin{equation}\label{N2b}
N_2(B\to 0)\propto\sqrt{\frac{B}{m}} \log^2\frac{B}{m}\to 0.
\end{equation}

This just reveals the nature of these states: The two-body valence contribution tends to zero, whereas the contributions containing the exchange particles dominate. Since the later many-body states, in addition to  exchange particles, can contain  valence ones, these states can be called  hybrid~states.

%%%%%%%%%%%%%%%%%%%%%%%%%%%
\section{Elastic Electromagnetic  and Transition Form Factors}\label{ffs}
The very different natures of the normal and abnormal states manifest themselves not only in their different contents (contributions of the valence and exchange particles),
but also  in the behaviors of their elastic electromagnetic form factors vs. the momentum transfer, as well as in the suppression  of the transition form factors 
(transitions: the normal $\to$ abnormal states) relative to the transitions between the states of the same nature.
The asymptotic the of elastic form factors is determined by the nature of states: A fast decrease is a manifestation of a many-body
structure \cite{matvmurtavk,brodsfarr,radyush}. 

We assume that one constituent particle is charged.
Knowing the BS amplitude, one can calculate the 
electromagnetic form factor of the system. For generality, we will  first consider the inelastic
transitions from the factor between the different states $i\to f$. To find the elastic one, we put $f=i$.

\begin{figure}[H]
%\vspace{0.8cm}
\includegraphics[width=7.5 cm]{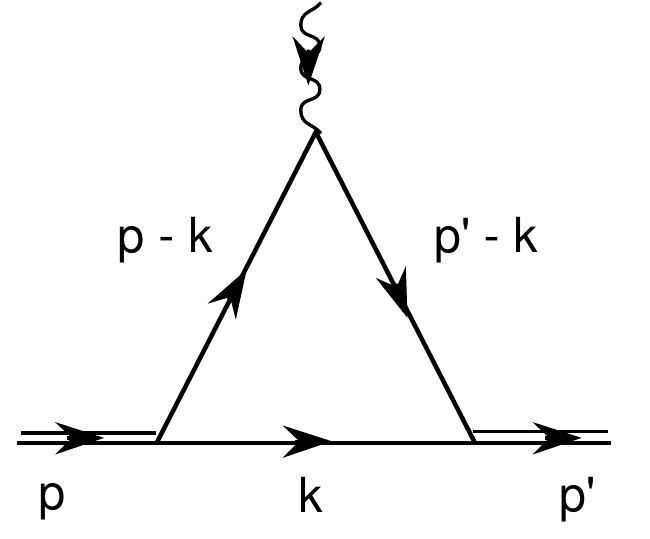}
\caption{Feynman diagram for the electromagnetic form factor. \label{triangle}}
\end{figure}

 First, we get the expression for the transition electromagnetic vertex $J_{\mu}$. It is presented in the Feynman graph in Figure~\ref{triangle}.
The left and right vertex functions in this diagram are expressed through the BS amplitude. In this way, we obtain the following expression for $J_{\mu}$ in terms of the BS amplitude (compare with  Equation (27) from \cite{cks21} and with Equation (7.1) from \cite{cdkm}):
\begin{equation}\label{ffbs}
J_{\mu}=i\int(p+p'-2k)_\mu 
\overline{\Phi}_f\left(\frac{1}{2}p'-k,p'\right)\; (k^2-m^2)\Phi_i \left(\frac{1}{2}p-k,p\right)\frac{d^4k}{(2\pi)^4}.                                                                                                                                                                                                                                                                                                                                                                                                                                
\end{equation}

It can be decomposed in terms of two covariant structures:%
%The scalar coefficients $F(Q^2)$ and $G(Q^2)$  in this decomposition are the form factors.

%In the case $M_f\neq M_i$, it has the following general decomposition in terms of two scalar
%functions:
%(see
%\footnote{We change the notations in comparison to Ref.
%\cite{tff}, where the form factor $G$ was denoted $F'$.}  \cite{tff}):

\begin{equation}\label{ffc}
J_{\mu}=\left[(p_{\mu}+{p'}_{\mu})+ ({p'}_{\mu}-p_{\mu})\frac{Q_c^2}
{Q^2}\right]F(Q^2)- ({p'}_{\mu}-p_{\mu})\frac{Q_c^2}{Q^2}G(Q^2).
\end{equation}

Here, $q=p'-p$,  $Q^2=-q^2=-(p'-p)^2$, and
%\begin{equation} \label{Qc2}
$Q_c^2={M_f}^2-M_i^2$,
%\end{equation}  
with $M_i$ and $M_f$ being the masses of the initial and final states,
respectively. The scalar coefficients $F(Q^2)$ and $G(Q^2)$ in this decomposition are the form factors that we intend to calculate.

From   Equation (\ref{ffc}), one can find the expressions of the form factors:
% it follows that the form factors are expressed through
%$J_{\mu}$ as
\begin{equation}\label{FFp}
F(Q^2)  = \frac{ (p+p')\cd J\,Q^2 + q\cd J \, Q_c^2}
{[(M_f-M_i)^2 +Q^2][(M_f+M_i)^2+Q^2]},\quad G(Q^2) = \frac{q\cd J}{Q_c^2}.
\end{equation}

From the conservation  of the electromagnetic current $J_{\mu}$, which is  expressed by the equality   $q\cd J=0$,  it follows that $G(Q^2) \equiv 0$ (for any $Q^2$). In Ref.  \cite{cks21}, {Appendix B}%MDPI: there is no appendix B in this article, please confirm and revise
, it is proven that this equality indeed follows from the BS equation. 
In other words, having found $\Phi(k,p)$ from the BS Equation~(\ref{bseq}), by substituting it into the current Equation (\ref{ffbs}) and extracting from it the form factor $G(Q^2)$ by means of Equation (\ref{FFp}), we obtain zero. Below, we present the elastic  form factors $F$ for a few normal and abnormal states, as well as the inelastic ones for transitions between them.

\begin{figure}[H]
%\vspace{0.8cm}
\includegraphics[width=10.5 cm]{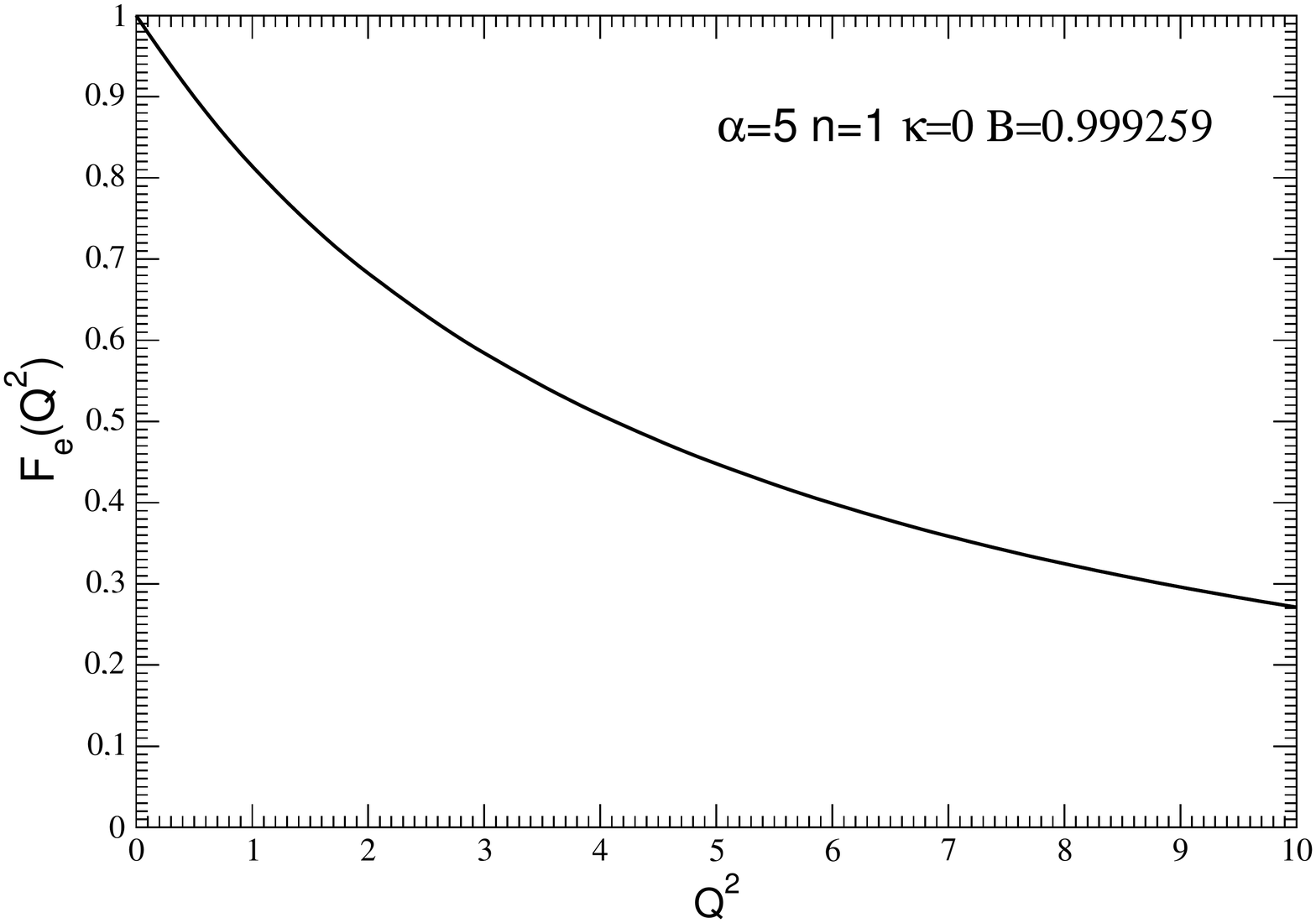}
\caption{{Elastic form factors} %MDPI: please change the comma in digits to dot (e.g., 0,1 should be 0.1)
 for the normal state with $n=1,\kappa=0$---No. 1 of Table \ref{tab1}. ({Adapted from} %MDPI: Please make sure that permission has been obtained and there is no copyright issue. 
 \protect{\cite{cks21}).}\label{Fel_1}}
\end{figure}

We start with the form factors of the states corresponding to $n=1$ and different $\kappa$s. The functions $g_{1\kappa}^0$ satisfy  Equation (\ref{gn}), 
where one should put $n=1$. The elastic form factor for the normal  state with  $n=1,\kappa=0$,  $B=0.999$ m---No. 1 of Table \ref{tab1}---is shown
in \mbox{Figure \ref{Fel_1}. }

\begin{figure}[H]
%\vspace{0.8cm}
\includegraphics[width=10.5 cm]{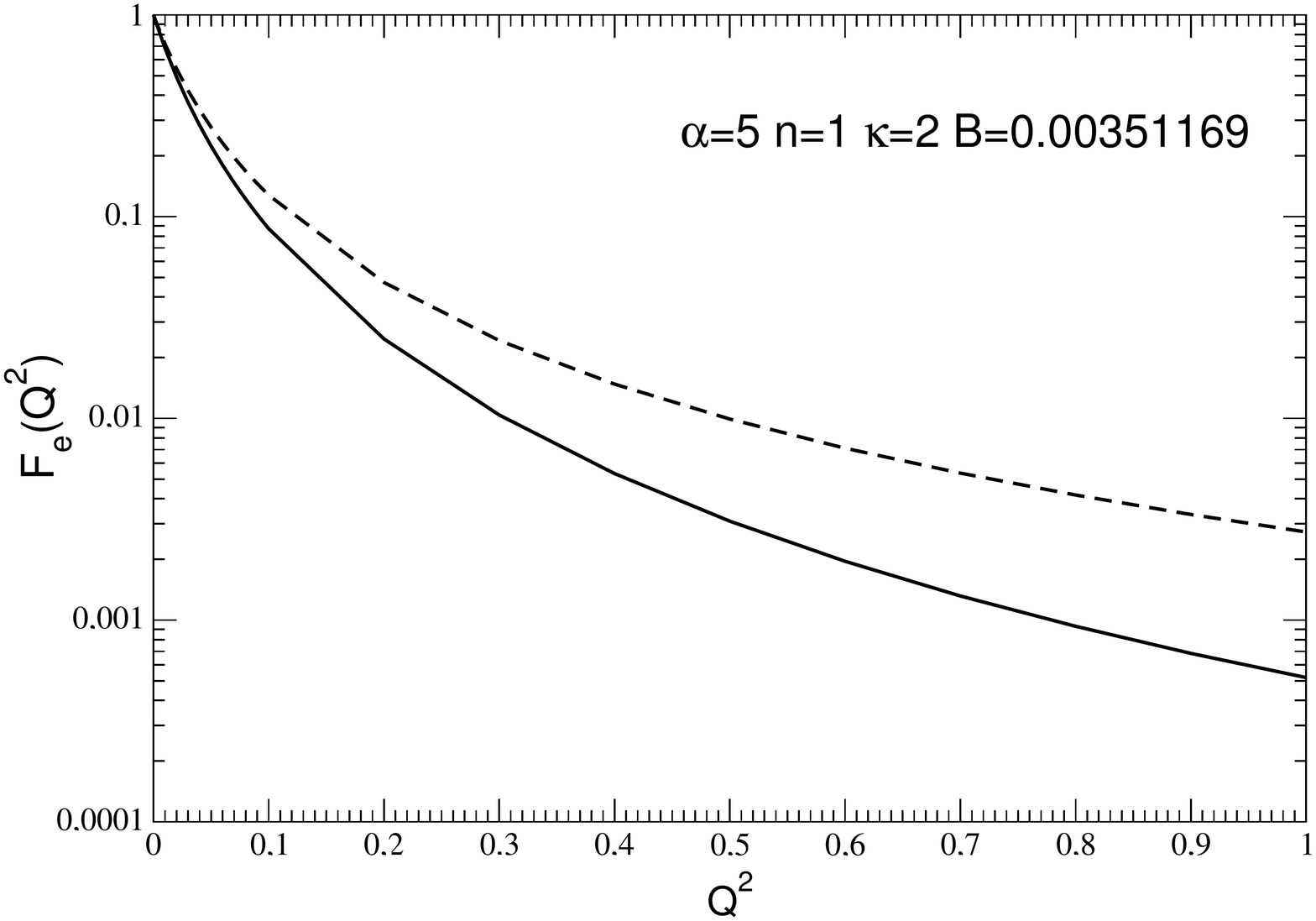}
\caption{{Solid line:} %MDPI: please change the comma in digits to dot (e.g., 0,1 should be 0.1)
  abnormal elastic form factor of  state No. 3 from  Table \ref{tab1} ($n=1,\kappa=2$).
The dashed line is the  normal elastic form factor of the state with $n=1,\kappa=0$  with the same
binding energy (and, hence, rms radius) as for the solid line. ({Adapted from} %MDPI: Please make sure that permission has been obtained and there is no copyright issue. 
 \protect{\cite{cks21}).}
\label{Fel_3}}
\end{figure}

The elastic form factors for the abnormal state with  $n=1,\kappa=2$,  $B=0.00351$ m---No. 3 of Table~\ref{tab1}---are shown
in Figure \ref{Fel_3} by the solid line. Comparing the solid curves in Figures \ref{Fel_1} and \ref{Fel_3}, we see that the abnormal-state elastic form factor decreases,  in the same interval of $0\leq Q^2\leq 1$, much faster (1000 times, approximately) than the normal one. There are a few reasons that are responsible for this faster decrease. Among them is the large size of the system due to the small binding energy, as well as its many-body content. In order to separate the many-body content,  we adjust the value of $\alpha$ of  state No. 1 so that it can have
the same binding energy as that of state No. 3. The result is shown in Figure \ref{Fel_3} by the dashed line. Though both curves have the same slope at the origin (that is, the systems have the same radii), the abnormal form factor still diminishes much faster than the normal one. At $Q^2=1$, the  abnormal/normal ratio is approximately 1/10. This confirms the many-body content of the abnormal states.

\begin{figure}[H]
%\vspace{0.8cm}
\includegraphics[width=10.5 cm]{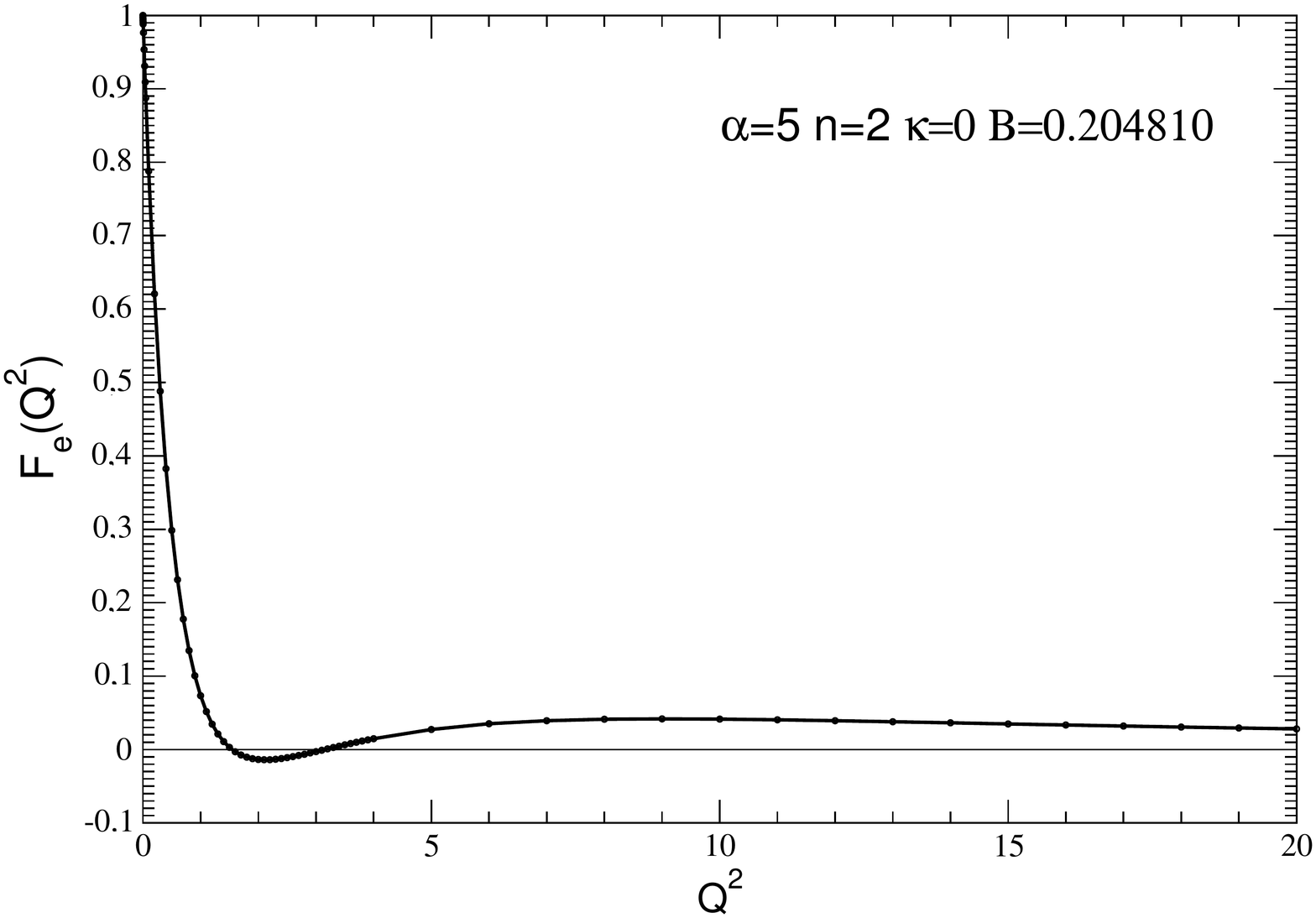}
\caption{{Elastic form} %MDPI: 1. please change hyphen (-) to minus sign. 2. please change the comma in digits to dot (e.g., 0,1 should be 0.1)
 factor of the  normal excited state with $n=2,\kappa=0$---No. 2 from Table \ref{tab1}. ({Adapted from} %MDPI: Please make sure that permission has been obtained and there is no copyright issue. 
 \protect{\cite{cks21}).}\label{Fel_2}}
\end{figure}

For the $n=2$ states, the situation is analogous. The elastic form factor 
of the normal excited state in No. 2 with $n=2$ , $\kappa=0$, and $B=0.2084$ m 
is shown in Figure \ref{Fel_2}. Though it corresponds to a comparable binding energy, it decreases much faster (by one order of magnitude at $Q^2 =1$)  than the form factor for the state with $n=1$ (No. 1), which is  shown in Figure \ref{Fel_1}. It has two zeroes and becomes negative in the interval $Q^2\in[1.5,3.0]$. This reflects the complex structure
of this system.

\begin{figure}[H]
%\vspace{0.8cm}
\includegraphics[width=10.5 cm]{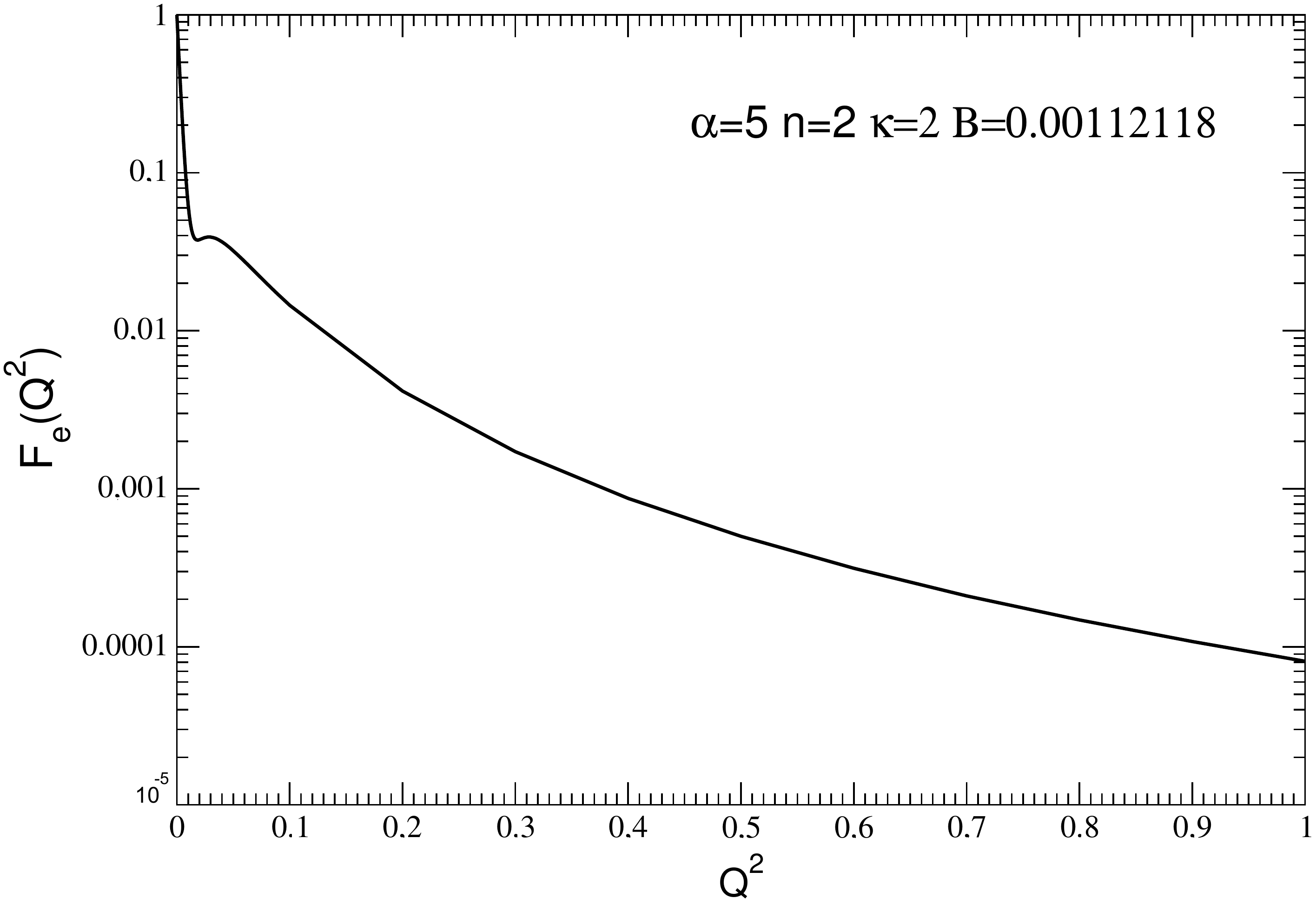}
\caption{\textls[-15]{{Elastic form} %MDPI: 1. Please use scientific notation. (e.g., 8 × 10³, not 8E3); 2. please change the comma in digits to dot (e.g., 0,1 should be 0.1)
 factors of the abnormal state $n=2, \kappa=2$---No. 4 of  Table \ref{tab1}. ({Adapted from}~%MDPI: Please make sure that permission has been obtained and there is no copyright issue. 
\protect{\cite{cks21}).}}\label{Fel_4}}
\end{figure}

Figure \ref{Fel_4} represents the elastic form factor of the abnormal state 
($n=2, \kappa=2, B=0.00112$~m---No. 4 in the Table \ref{tab1}). 
It also decreases faster than  the $n=1$, $\kappa=2$ state
 (No. 3) with a binding energy of the same order. Let us also emphasize  the irregularity, which is  similar to that of a diffraction structure.
 Irregularities such as this one are normally absent in the ground-state form factors of two-body scalar systems. 
Their manifestation in the form factors of the excited state with $n>1$---both normal and abnormal---shows the complexity of these systems. 

\begin{figure}[H]
%\vspace{0.8cm}
\includegraphics[width=10.5 cm]{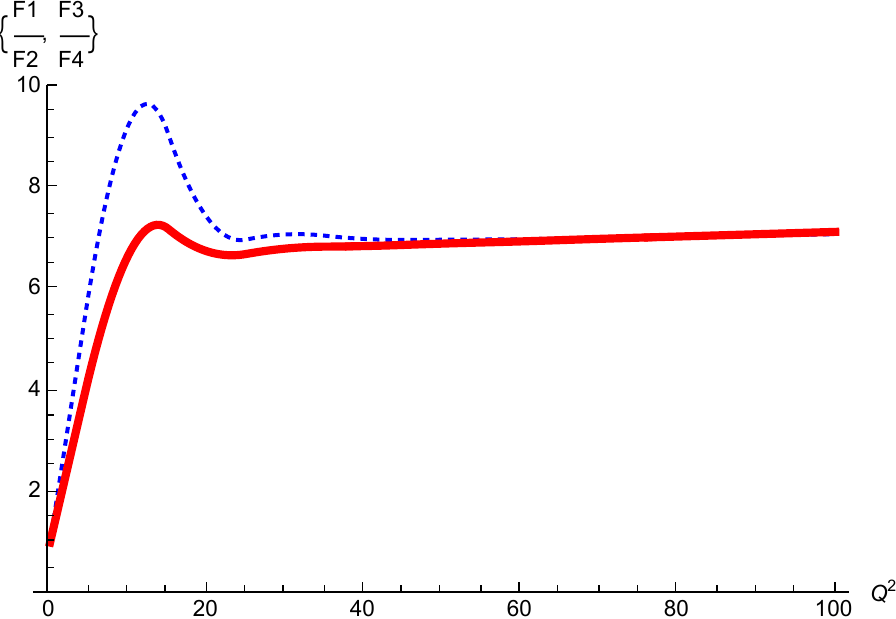}
\caption{The ratios of the form factors for the states with $n=1$ and $n=2$. The dotted curve is the ratio for  states Nos. 1 and 2 from  Table \ref{tab1}, 
$\kappa=0$. The solid curve corresponds to the ratio for  states Nos. 3 and 4, $\kappa=2$. ({Adapted from} %MDPI: Please make sure that permission has been obtained and there is no copyright issue. 
 \protect{\cite{cks21}).} \label{rat1234}}
\end{figure}

To reveal the influence of the quantum number $n$ on the form factors and on their asymptotic,   in Figure \ref{rat1234}, we show the ratios of the form factors for the states with $n=1$ and $n=2$ and for the same fixed $\kappa$, either $\kappa=0$ or 2. At large values of $Q^2$, these ratios tend to the constant, which is the same (equal to $\sim$7) for both ratios.

\begin{figure}[H]
%\vspace{0.8cm}
\includegraphics[width=10.5 cm]{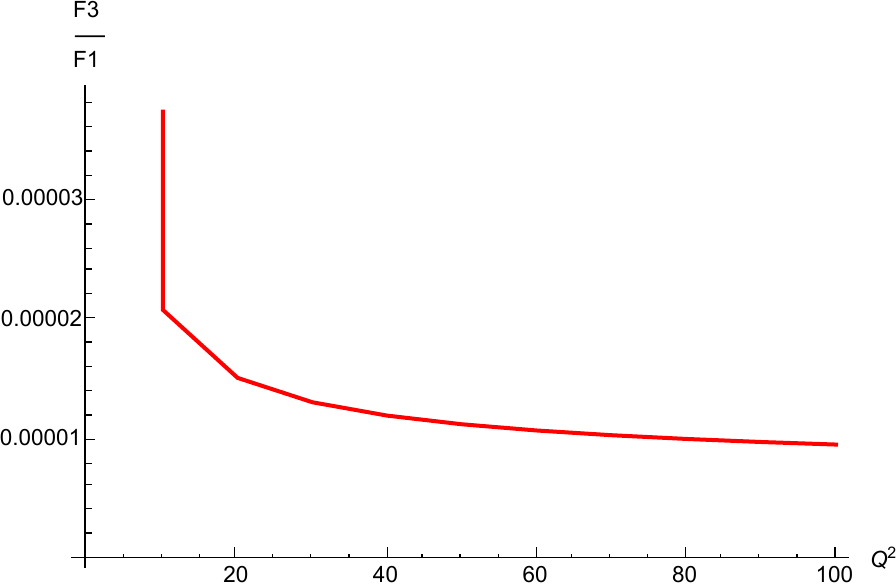}
\caption{The form factor ratio:  abnormal
state No. 3/normal state No. 1 from Table \ref{tab1}. ({Adapted from} %MDPI: Please make sure that permission has been obtained and there is no copyright issue. 
 \protect{\cite{cks21}).} \label{rat31}}
\end{figure}

Note that the ratio of the abnormal/normal form factors  also tends to be constant at large values of $Q^2$; however, this constant is much smaller than the normal/normal and abnormal/abnormal ratios. This is seen in the comparison of Figure \ref{rat1234} with Figure \ref{rat31}. The latter figure shows the ratio of the  abnormal/normal form factors (states Nos. 3 and  1). The value of the constant for the asymptotic of this ratio is $\sim 10^{-5}$.  

Since the form factors decrease at different speeds for systems with different numbers of constituents, the fact that all of the form factor ratios shown in Figures \ref{rat1234}
and \ref{rat31} tend to be constants, though they are different for different states, supports the interpretation of the abnormal states as hybrid systems that contain both  constituent (in small amounts) and exchange particles (dominant). Then, the asymptotic of the form factor is always determined by the  two-body contributions, which are, however, very different for normal and abnormal states and are very small for the latter.

\begin{figure}[H]
\hspace{3cm}
\includegraphics[width=5 cm]{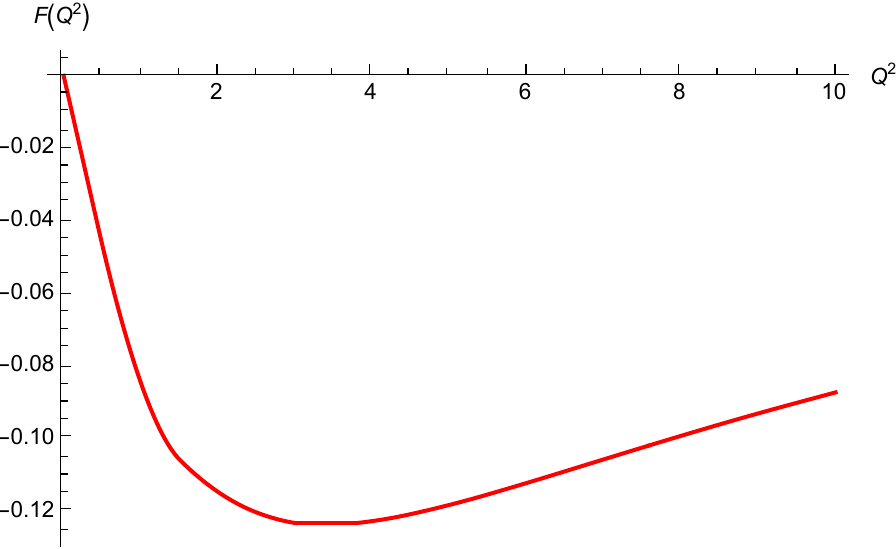}\\

\vspace{0.5cm}
%\newline
\includegraphics[width=5 cm]{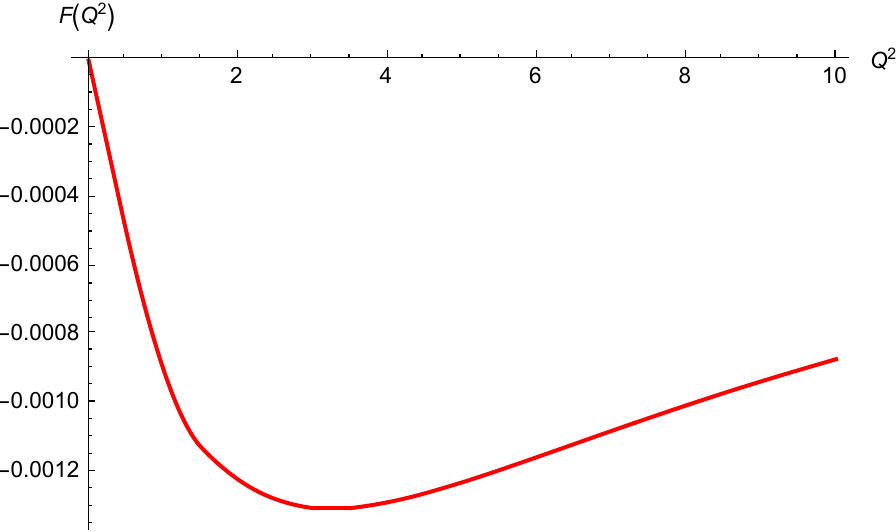}
\hspace{1cm}
\includegraphics[width=5 cm]{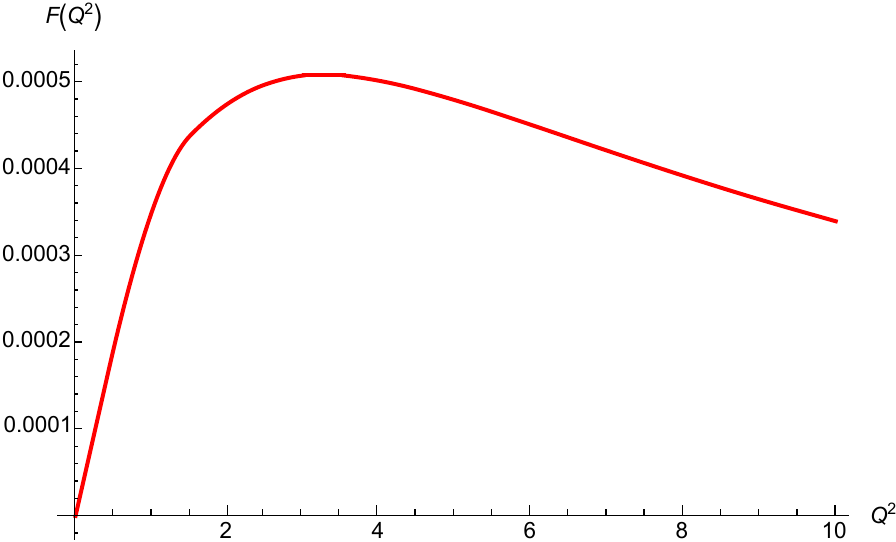}
\caption{Form factors for the
transitions from   state  No. 1 
%($n=1,\kappa=0$) 
to the states with $\kappa=0,2$ listed in Table~\ref{tab1}. ({Adapted from} %MDPI: Please make sure that permission has been obtained and there is no copyright issue. 
 \cite{cks21}).
Upper panel: normal (No. 1, $n=1,\kappa=0$) $\to$   normal (No. 2, $n=2, \kappa=0$). 
Lower left panel: normal (No. 1, $n=1,\kappa=0$) $\to$ abnormal (No. 3, $n=1, \kappa=2$). 
Lower right panel: normal (No. 1, $n=1,\kappa=0$) $\to$  abnormal (No. 4,  $n=2, \kappa=2$).}
\label{Ftr1_234}
%\end{center}
%\end{minipage}
\end{figure}

\begin{figure}[H]
\hspace{3cm}
\includegraphics[width=5 cm]{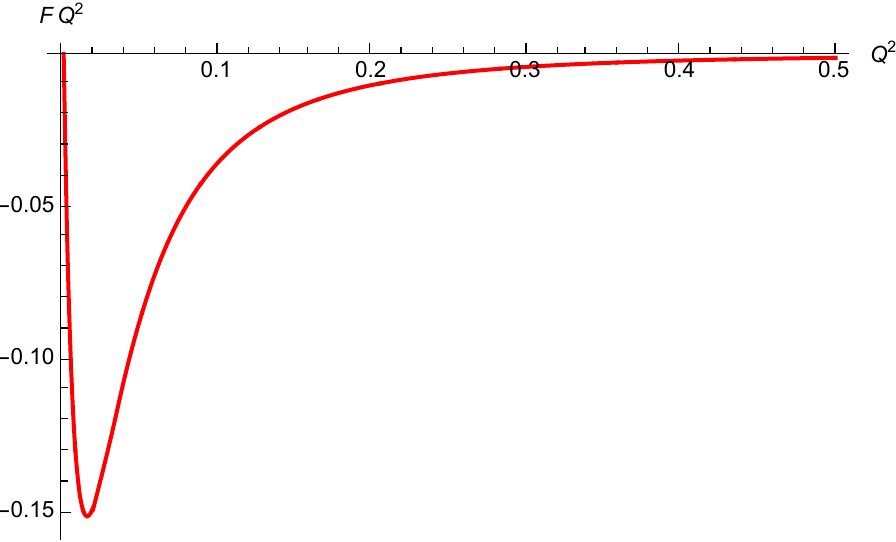}\\

\includegraphics[width=5 cm]{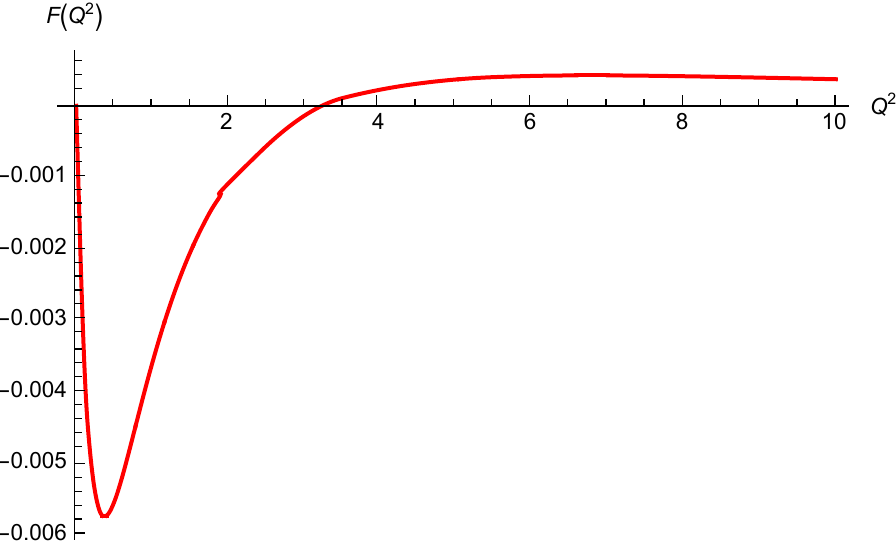}
\hspace{1cm}
\includegraphics[width=5 cm]{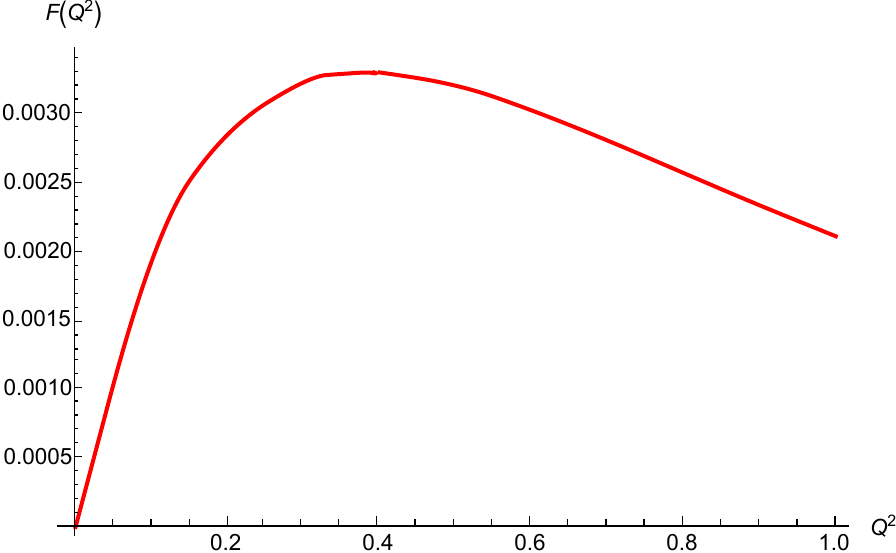}
\caption{Transition form factors between the following states taken from Table \ref{tab1}.  ({Adapted from}%MDPI: Please make sure that permission has been obtained and there is no copyright issue. 
~\cite{cks21}). 
Upper panel: abnormal (No. 3, $n=1,\kappa=2$) $\to$ abnormal (No. 4, $n=2,\kappa=2$). 
Lower left panel: normal (No. 2, $n=2,\kappa=0$) $\to$ abnormal (No. 3, $n=1,\kappa=2$). 
Lower right panel: normal (No. 2, $n=2, \kappa=0$) $\to$ abnormal (No. 4, $n=2,\kappa=2$).}
\label{F34_23_24}
\end{figure}

So far, we discussed the elastic form factors. The transition form factors are also very informative for understanding the structures of the abnormal states. We restrict ourselves to four states with $n=1,2;\kappa=0,2$ ({Nos.} % Should this be a dash instead of a \div sign?
 $1\div 4$ in the Table \ref{tab1}). There are six transitions between four states. Figures \ref{Ftr1_234} and \ref{F34_23_24} represent all  six of these transition form factors. 

One can see that the transition form factors between the states of the same nature---normal--normal and abnormal--abnormal---are much larger than for the transitions between the states of different nature (normal--abnormal). Thus, the transition form factor between two normal states---No. 1
($n=1,\kappa=0$) and No. 2 ($n=2,\kappa=0$), which are shown in the upper panel of Figure \ref{Ftr1_234}---is 
 larger than  the maximal values of the transition  form factors  from the normal to the 
abnormal states, which are  shown in the lower panel, by a factor of~$\sim$100.

The same property is confirmed by Figure \ref{F34_23_24}.
The maximal value of the transition form factor between two abnormal states---No. 3
($n=1,\kappa=2$) and No. 4 ($n=2,\kappa=2$), which are shown in the  upper panel of Figure  \ref{F34_23_24}---has the same
order of magnitude as the normal--normal one shown in the  upper panel of Figure
\ref{Ftr1_234}. However, it decreases much faster with the increase in $Q^2$. At the same time, this form factor is again 
larger than  the maximal values of the transition  form factors  from the normal to the
abnormal states, which are shown in the lower panels  of Figures  \ref{Ftr1_234} and \ref{F34_23_24}, by a factor of $\sim$100. These relations are apparently caused by the need to rebuild the  structure of the system when the transition from a normal to an abnormal state takes place.

Let us now come back to Equation (\ref{ffc}) for the electromagnetic current. For the transitions, it contains an extra form factor $G(Q^2)$.
This is determined by Equation  (\ref{FFp}). However, as mentioned, from the current conservation, it follows that $G(Q^2)\equiv 0$ for any $Q^2$. A numerical check of this
equality provides a strong test of the numerical calculations carried out in \cite{cks21}. Therefore, together with  the transition form factors presented above,
 the transition form factors $G(Q^2)$  were also calculated for all transitions.

\begin{figure}[H]
%\vspace{0.8cm}
\includegraphics[width=10.5 cm]{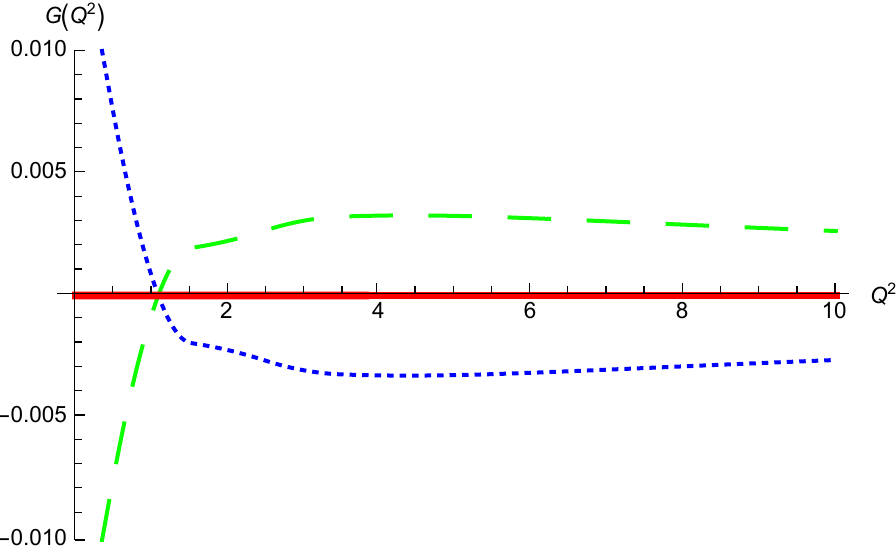}\\
\caption{Contributions to the
form factor $G(Q^2)$ of the transition between the states Nos. 2 and 3 from  Table \ref{tab1}.
The dotted line is $G^{00}(Q^2)$, and the dashed line is   $G^{10}(Q^2)$ (for their definitions, see  the text). 
The their sum   (solid line) is the full form factor $G(Q^2)$. ({Adapted from} %MDPI: Please make sure that permission has been obtained and there is no copyright issue. 
 \protect{\cite{cks21}).}
\label{G_23}}
\end{figure}

The results are illustrated in Figure \ref{G_23} with the example of the transition between   state
No.~2 $(n=2,\kappa=0)$ and state No. 3 $(n=1,\kappa=2)$.
For the latter state, the BS amplitude is given by Equation (\ref{Phi1}), which contains one function, $g_1^0(z)$.
For  state No. 2,  the BS amplitude is given by \mbox{Equation (\ref{Phi2})}, which contains a sum that includes two functions, $g_2^0(z)$
and $g_2^1(z)$. Therefore, the form factor $G(Q^2)$ is determined by the sum of two terms:  $G^{00}(Q^2)$ (proportional to $g_2^0g_1^0$)
and $G^{10}(Q^2)$  (proportional to $g_2^1g_1^0$). They are shown
in Figure \ref{G_23} by dotted ($G^{00}(Q^2)$) and dashed ($G^{10}(Q^2)$) lines, respectively. Their sum, that is, the full transition form
factor $G(Q^2)\approx 10^{-6}$, is given by a thick solid line and is 
indistinguishable from zero on the scale of the figure. So, this test, which is extremely sensitive to any inaccuracy, is successfully satisfied.
For example, a relative error in the calculation of  the binding energy of the order of $\sim 10^{-4}$ destroys the cancellation seen in Figure \ref{G_23}. The value of 
$G(Q^2)$ takes on  the order of the dashed and dotted curves.

The results of the calculations shown in Figures \ref{Fel_1} and \ref{Fel_2} for the normal states, as well as their comparisons with Figures  \ref{Fel_3} and \ref{Fel_4}
for the elastic abnormal form factors, clearly indicate that, with respect to  $Q^2$, the latter ones diminish considerably faster than the normal form factors.
This affirms that  abnormal bound systems are dominated by  many-body Fock sectors \cite{matvmurtavk,brodsfarr,radyush}. 

The normal--abnormal transitions  are considerably suppressed  in comparison
to the normal--normal and abnormal--abnormal ones. 
This is a manifestation of the fact that the normal and abnormal
states have different levels of compositeness. Therefore, the normal--abnormal transitions entail the 
rebuilding of these states.

%The current conservation becomes a stringent self-consistency criterion of our results: the calculated form factor $G(Q^2)$ should
%be identically zero, or very small  within the numerical uncertainties.

 %and find in this way the contribution of these two constituents to the system. Therefore, the BS amplitude (\ref{bs}) and corresponding equation are a %powerful and convenient tool for studying the relativistic systems. As mentioned, for the solution of the first type $N_2=1$, the spectrum and the wave %function coincide with the non-relativistic ones. 

%%%%%%%%%%%%%%%%%%%%%%%%%%%%%%%%%%%%%%%%%%
\section{Discussion}\label{discuss}

Our consideration is based on  a conception that is widely used in nuclear and particle physics: {Massive particles (constituents, valence particles) interact via exchanges with other light particles.} % Please confirm that meaning has been retained
 We assume that the exchanges are massless. Traditionally, this model is applied to  bound systems made of  constituents. Indeed, in a static approximation, the interaction is reduced to the potential, and we deal with a non-relativistic system with a fixed number of constituents, as described by the Schr\"odinger equation.

 We started this article with physical arguments supporting the existence of  bound systems of quite different natures---relativistic systems dominated by    exchange particles.  
 There are two arguments (at least) in favor of this hypothesis: ({\it i})~the non-perturbative nature of  bound states, which implies  multiple exchanges;
  ({\it ii})~the manifestation of  relativistic retardation effects in the propagation of  exchange particles
beyond the narrow domain of validity of the static approximation. {These effects result in the filling of the intermediate states by the exchange particles, that is, in the appearance and domination in the state vector of the corresponding Fock sectors.} % Please confirm that meaning has been retained
 Since two limiting cases (non-relativistic and ultra-relativistic) are associated with one and the same initial field-theoretical  Hamiltonian,  one can expect that they---if they exist---can be found from one and the same equation in its non-relativistic and fully relativistic domains. 
  
This is exactly what was found by Wick and Cutkosky in their model \cite{wick, cutk}. For massless exchange, in the non-relativistic limit, they reproduced the 
ordinary Coulomb spectrum and the wave functions. In addition, they  found new states that had no non-relativistic counterparts and that disappeared in the non-relativistic limit. The physical meaning of these new states remained unclear {and controversial} for a long time. % Please confirm that meaning has been retained

\textls[-15]{By analyzing the Fock sector content  of both normal and abnormal states, we  \mbox{found \cite{cks21,hvk2004}}} that the states of these two types drastically differ from each other. 
The normal states are dominated by  constituent particles (as expected). On the contrary, the constituent contribution to the abnormal states is small. The latter  have a different nature---they are dominated by  exchange particles. The method that we use gives  information about the probability of a two-body contribution, which is very small (see $N_2$ in  Table \ref{tab1}, lines $3\div 6$), as well as information about the sum of the other ones (which dominate). This sum contains  the contributions of ``two constituents plus one exchange particle'' + ``two constituents plus two exchange particles'', etc. However, we do not know the probability  of each of these extra Fock sectors. {However, the qualitative physical picture discussed in the introduction and the  
fast decrease in the elastic electromagnetic form factors} (Section \ref{ffs}) {speak in favor of the dominance of the Fock sectors of ``two constituents plus many exchange particles''.} % Please confirm that meaning has been retained
 This compositeness is also confirmed by a comparison of the values and behaviors of the form factors for the normal$\to$normal, abnormal$\to$abnormal, and normal$\leftrightarrow$abnormal transitions. The latter transitions require the rebuilding of the states, and therefore, they are especially~suppressed. 

For these reasons, we conclude that the abnormal states found in the BS framework correspond to bound systems of a relativistic origin that are dominated by  
massless exchange particles. They do not indicate  a  pathology in the BS equation (contrary to what is sometimes supposed in the literature). 
  
The essentially non-perturbative origin of the abnormal states raises the question of the validity of the ladder approximation. The more important question is: Though a solvable model implies a simplified (ladder) kernel,  are the abnormal states just a consequence of  the ladder approximation?
The multiparticle contributions to the BS kernel and their influence on  the abnormal states have not been investigated. Since  multiparticle Feynman graphs add extra particles in the intermediate states, one can expect that they increase the contributions of higher Fock sectors. It should at least be noted that  the cross-ladder contribution to the ladder kernel increases the binding energy \cite{ck_bs2}. It increases the relativistic effects in the system and, therefore, should increase the many-body contributions. Therefore, it works in favor of, not against, the existence of the abnormal states.
 
In any case, although the Wick--Cutkosky model is oversimplified, it contains the phenomenon of  intermediate particle creation, which is crucial for generating  multiparticle Fock sectors.
There are no  indications that this generation is an artifact of  ladder approximation. On the contrary, the generation of  exchange particles in a bound system is a consequence of their creation in the  intermediate kernel states and their manifestation in a relativistic system due to retardation.
This phenomenon, even in the ladder kernel framework, provides an example of the natural formation of  hybrid states.
Therefore, one can expect that more complicated kernels and more sophisticated field theories also result in the formation of the states of this kind.
However, since the binding energies of these systems are extremely small, the methods of their detection and the question of the ``smoking gun'' deserve special study.

The systems dominated by  exchange particles can have an electromagnetic nature. However, 
in view of the results discussed above, we also mention glueballs and hybrid states predicted in QCD (for a review, see \cite{klempt,vento,ochs} and the references therein). In principle,    glueballs originate for different reasons, such as self-interactions of gluons. They do not contain  constituent quarks. However, the states considered in the present article have a hybrid nature---they appear due to  exchanges between constituents. However, as we saw, the contribution of the Fock component containing the constituents only  tends to zero for small binding energies. 
%However, the contribution of the sector with two constituents plus exchange particles can exist - we know nothing about its numerical value. 
In addition, we considered scalar massless colorless exchanges, where the problem of the color compositeness of  hybrid states does not appear. However,  glueballs made of  colored gluons must be colorless. This imposes restrictions on their compositeness. In our opinion,  these differences, however, can be considered as secondary. Generally, the main reason for the origination of systems dominated by massless particles is the 
possibility of easy virtual creation of the latter  in the intermediate states, independently of the particular mechanisms of their creation:  self-interaction or exchanges between constituents in the ladder approximation or beyond it. From this general point of view, the states discussed in the present article and  glueballs can be considered as being akin to each other.

So far, we discussed the results found in the case of massless exchange particles.  Research on the case of massive exchange particles is in progress.

%Authors should discuss the results and how they can be interpreted from the perspective of previous studies and of the working hypotheses. The findings %and their implications should be discussed in the broadest context possible. Future research directions may also be highlighted.

%%%%%%%%%%%%%%%%%%%%%%%%%%%%%%%%%%%%%%%%%%
%\section{Conclusions}

%This section is not mandatory, but can be added to the manuscript if the discussion is unusually long or complex.

%%%%%%%%%%%%%%%%%%%%%%%%%%%%%%%%%%%%%%%%%%
%\section{Patents}
%This section is not mandatory, but may be added if there are patents resulting from the work reported in this manuscript.

%%%%%%%%%%%%%%%%%%%%%%%%%%%%%%%%%%%%%%%%%%
\vspace{6pt} 

%%%%%%%%%%%%%%%%%%%%%%%%%%%%%%%%%%%%%%%%%%
%% optional
%\supplementary{The following are available online at \linksupplementary{s1}, Figure S1: title, Table S1: title, Video S1: title.}


%%%%%%%%%%%%%%%%%%%%%%%%%%%%%%%%%%%%%%%%%%
%\authorcontributions{For research articles with several authors, a short paragraph specifying their individual contributions must be provided. The following %statements should be used ``Conceptualization, X.X. and Y.Y.; methodology, X.X.; software, X.X.; validation, X.X., Y.Y. and Z.Z.; formal analysis, X.X.; %investigation, X.X.; resources, X.X.; data curation, X.X.; writing---original draft preparation, X.X.; writing---review and editing, X.X.; visualization, X.X.; %supervision, X.X.; project administration, X.X.; funding acquisition, Y.Y. All authors have read and agreed to the published version of the manuscript.'', %please turn to the  \href{http://img.mdpi.org/data/contributor-role-instruction.pdf}{CRediT taxonomy} for the term explanation. Authorship must be limited to %those who have contributed substantially to the work~reported.}

\funding{This research received no external funding.}

\conflictsofinterest{The author declares no conflict of interest.} 

%% Optional
%\sampleavailability{Samples of the compounds ... are available from the authors.}

%%%%%%%%%%%%%%%%%%%%%%%%%%%%%%%%%%%%%%%%%%
%% Only for journal Encyclopedia
%\entrylink{The Link to this entry published on the encyclopedia platform.}

%%%%%%%%%%%%%%%%%%%%%%%%%%%%%%%%%%%%%%%%%%
%% Optional
\clearpage
\abbreviations{Abbreviations}{
The following abbreviations are used in this manuscript:\\

\noindent 
\begin{tabular}{@{}ll}
BS & Bethe--Salpeter\\
LF & Light front\\
%MDPI & Multidisciplinary Digital Publishing Institute\\
%DOAJ & Directory of open access journals\\
%TLA & Three letter acronym\\
%LD & Linear dichroism
\end{tabular}}

%%%%%%%%%%%%%%%%%%%%%%%%%%%%%%%%%%%%%%%%%%
%% Optional
\appendixtitles{yes} % Leave argument "no" if all appendix headings stay EMPTY (then no dot is printed after "Appendix A"). If the appendix sections contain a heading then change the argument to "yes".
\appendixstart
\appendix
\section{Calculating \boldmath{$N_{tot}$} in the Limit \boldmath{$B\to 0$}}\label{app1}
For any given state, the expression for $N_{tot}$ at $B\to 0$ in terms of the solution $g(z)$ corresponding to this state is found from the condition $F_{el}(0)=1$,  where $F_{el}(Q)$ is the elastic form factor of this state. It is determined by Equation (A.11) from \cite{cks21}:
\begin{equation}\label{Ntot1}
N_{tot}(B\to 0)=\frac{3m^{5/2}}{2^6\pi B^{5/2}}
\int_0^1 dz'\int_0^1dz\frac{z^2{z'}^2 g(z)g(z')}{(z+z')^5}.
\end{equation}

For the normal $n=1$ state, one has $g(z)=1-|z|$ \cite{wick,cutk,nak69}. Substituting this into (\ref{Ntot1}) and integrating over $z,z'$, we find:
\begin{equation}\label{Ntot2}
N_{tot}(B\to 0)=\frac{m^{5/2}}{2^{10}\pi B^{5/2}}.
\end{equation}

The substitution of this expression into  (\ref{norm10}) results in $N_2(B\to 0)=1$  with the correction determined by Equation (\ref{N2a}).

For  abnormal states with $\kappa\ge 1$---still in the limit $B\to 0$---the solution reads \cite{wick,cutk,nak69}:
\begin{equation}\label{sol}
g_{\kappa n}(z)=(1-z^2)^n |z|^{\frac{1}{2}+\rho}F\left(\frac{1}{2}\Bigl(\frac{3}{2}+\rho+n\Bigr),\frac{1}{2}\Bigl(\frac{1}{2}+\rho+n\Bigr),n+1;1-z^2\right),
\end{equation}
where $F$ is the hypergeometric function, $\lambda=\frac{\alpha}{\pi}$,
%\begin{equation}\label{param}
$$
\rho=\sqrt{\frac{1}{4}-\lambda},\quad \lambda=\frac{1}{4}+\frac{\pi^2(\kappa-1)^2}{\left[\log\left(1-\frac{1}{4}M^2\right)\right]^2}.
$$
%\end{equation}

The two-body contribution is given by Equation (\ref{norm10}). It contains  $Q^3\approx\left(z^2+\frac{B}{m}\right)^3$ in the denominator. Therefore, the integral 
(\ref{norm10}) for $N_2$ is determined by the domain $z\to 0$. In this domain and  for $\rho\to 0$, the solution (\ref{sol}) for $n=1,\kappa=2$ takes the form:
$$
g_{21}(z)\approx -\frac{4 \sqrt{2 z} }{\pi} \left(2 + \log\frac{z}{8}\right).
$$

Then, by calculating $N_2$, we find for the leading term:
\begin{equation}\label{N2al}
N_2(B\to 0)\approx  \frac{\log^2\frac{B}{m}}{96 B^2 N_{tot}}
\end{equation}

$N_{tot}$ is given by Equation (\ref{Ntot1}), where $g(z)$ is determined by Equation (\ref{sol}) with \mbox{$n=1,\kappa=2$}, $\rho\to 0$. The dependence on $B$ is determined by the factor $1/B^{5/2}$, whereas, in this case, the double integral cannot be calculated analytically, but is calculated numerically: $N_{tot}(B\to 0)\approx 8\times 10^{-4}(m/B)^{5/2}$. By substituting this expression into Equation (\ref{N2al}), we obtain Equation (\ref{N2b}) for the abnormal $N_2(B\to 0)$.
%%%%%%%%%%%%%%%%%%%%%%%%%%%%%%%%%%%%%%%%%%
%\end{paracol}
%%%%%%%%%%%%%%%%%%%%%%%%%%%%%%%%%%%%%%%%%%
% To add notes in main text, please use \endnote{} and un-comment the codes below.
%\begin{adjustwidth}{-5.0cm}{0cm}
%\printendnotes[custom]
%\end{adjustwidth}
%%%%%%%%%%%%%%%%%%%%%%%%%%%%%%%%%%%%%%%%%%

%\emph{To add: 1) References to our previous papers. 2) Discussion of the "intermediate" situation, when the constituent 
%and exchange particles are 50-50 \%.
%3) To mention massive case.}

\begin{adjustwidth}{-\extralength}{0cm}
%\centering %% If there is a figure in wide page, please release command \centering
\reftitle{References}
\printendnotes[custom]

%\externalbibliography{yes}
%\bibliography{your_external_BibTeX_file}

%=====================================
% References, variant B: internal bibliography
%=====================================

\end{adjustwidth}

% If authors have biography, please use the format below
%\section*{Short Biography of Authors}
%\bio
%{\raisebox{-0.35cm}{\includegraphics[width=3.5cm,height=5.3cm,clip,keepaspectratio]{Definitions/author1.pdf}}}
%{\textbf{Firstname Lastname} Biography of first author}
%
%\bio
%{\raisebox{-0.35cm}{\includegraphics[width=3.5cm,height=5.3cm,clip,keepaspectratio]{Definitions/author2.jpg}}}
%{\textbf{Firstname Lastname} Biography of second author}

% The following MDPI journals use author-date citation: Admsci,  Arts, Econometrics, Economies, Genealogy, Humanities, IJFS, Jintelligence, JRFM, Languages, Laws, Literature, Religions, Risks, Social Sciences. For those journals, please follow the formatting guidelines on http://www.mdpi.com/authors/references
% To cite two works by the same author: \citeauthor{ref-journal-1a} (\citeyear{ref-journal-1a}, \citeyear{ref-journal-1b}). This produces: Whittaker (1967, 1975)
% To cite two works by the same author with specific pages: \citeauthor{ref-journal-3a} (\citeyear{ref-journal-3a}, p. 328; \citeyear{ref-journal-3b}, p.475). This produces: Wong (1999, p. 328; 2000, p. 475)

%%%%%%%%%%%%%%%%%%%%%%%%%%%%%%%%%%%%%%%%%%
%% for journal Sci
%\reviewreports{\\
%Reviewer 1 comments and authors’ response\\
%Reviewer 2 comments and authors’ response\\
%Reviewer 3 comments and authors’ response
%}
%%%%%%%%%%%%%%%%%%%%%%%%%%%%%%%%%%%%%%%%%%
\end{document}